\documentclass[10pt,oneside,onecolumn]{elsart}

\usepackage[ansinew]{inputenc}
\usepackage[english]{babel}
\usepackage{ifthen,shortvrb}
\usepackage[fleqn]{amsmath}
\usepackage{amsxtra,amsfonts,latexsym,amssymb,euscript}
\usepackage{amstext}
\usepackage{graphicx}
\usepackage{dcolumn}
\usepackage{subfigure}
\usepackage{natbib}
\usepackage{verbatim}
\usepackage[hidelinks]{hyperref}
\usepackage{color}

\newcommand{\bx}{\mathbf{x}}
\newcommand{\bt}{\mathbf{t}}

\newcommand{\by}{\mathbf{y}}
\newcommand{\bw}{\mathbf{w}}
\newcommand{\bbu}{\bar{\mathbf{u}}}
\newcommand{\bp}{\mathbf{p}}
\newcommand{\bQ}{\mathbf{H}}
\newcommand{\bC}{\mathbf{C}}
\newcommand{\bu}{\mathbf{u}}
\newcommand{\bM}{\mathbf{M}}
\newcommand{\bI}{\mathbf{I}}
\newcommand{\IC}{\mathcal{I}}
\newcommand{\II}{\mathcal{I}_0}
\newcommand{\bs}{\mathbf{s}}

\newcommand{\bb}{\mathbf{b}}
\newcommand{\bbw}{\bar{\mathbf{w}}}
\newcommand{\bz}{\mathbf{0}}

\newcommand{\eqdef}{\triangleq}

\newcommand{\rr}{\mathbb{R}}
\newcommand{\matrice}[2]{\left[\hspace*{-.1cm}\ba{#1} #2 \ea\hspace*{-.1cm}\right]}
\newcommand{\ba}[1]{\begin{array}{#1}}
\newcommand{\ea}{\end{array}}
\newcommand{\st}{{\rm s.t.}}

\newtheorem{problem}{Problem}
\newtheorem{property}{Property}
\newtheorem{example}{Example}

\begin{document}

\begin{frontmatter}
\date{01/11/2014}
\journal{Int. J. Solids and Structures}
\title{Optimization algorithms for the solution of the frictionless normal contact between rough surfaces}
\author{A. Bemporad} and \author{M. Paggi\corauthref{mp}}
\address{IMT Institute for Advanced Studies Lucca,\\ Piazza San Francesco 19, 55100 Lucca, Italy}
\corauth[mp]{Corresponding author. Tel: +39-0583-4326-604, Fax:
+39-0583-4326-565.} \ead{marco.paggi@imtlucca.it}
\begin{abstract}
This paper revisits the fundamental equations for the solution of
the frictionless unilateral normal contact problem between a rough
rigid surface and a linear elastic half-plane using the boundary
element method (BEM). After recasting the resulting Linear
Complementarity Problem (LCP) as a convex quadratic program (QP)
with nonnegative constraints, different optimization algorithms are
compared for its solution: ($i$) a Greedy method, based on different
solvers for the unconstrained linear system (Conjugate Gradient CG,
Gauss-Seidel, Cholesky factorization), ($ii$) a constrained CG
algorithm, ($iii$) the Alternating Direction Method of Multipliers
(ADMM), and ($iv$) the Non-Negative Least Squares (NNLS) algorithm,
possibly warm-started by accelerated gradient projection steps or
taking advantage of a loading history. The latter method is two
orders of magnitude faster than the Greedy CG method and one order
of magnitude faster than the constrained CG algorithm. Finally, we
propose another type of warm start based on a refined criterion for
the identification of the initial trial contact domain that can be
used in conjunction with all the previous optimization algorithms.
This method, called Cascade Multi-Resolution (CMR), takes advantage
of physical considerations regarding the scaling of the contact
predictions by changing the surface resolution. The method is very
efficient and accurate when applied to real or numerically generated
rough surfaces, provided that their power spectral density function
is of power-law type, as in case of self-similar fractal surfaces.
\end{abstract}
\end{frontmatter}

\section{Introduction}
\label{sec:intro} Contact mechanics between rough surfaces is a very
active area of research for its paramount importance to address
several practical applications in physics and engineering.
Understanding the evolution of the contact domain and contact
variables, such as load, real contact area, contact stiffness, and
many others, that depend on the morphological properties of
roughness, is still considered a challenging problem today. The
reader is referred to
\citep{bounds,noson,Ciavarella1,HR,Ciavarella2,Ciavarella3,CB08,PC10,CPM,PB,PPP,yastrebov}
for an overview of research results developed during the last
decade.

Semi-analytical contact theories that are able to provide synthetic
predictions of the contact response is also a challenging topic. A
comparison and validation on benchmark results is necessary to
understand the limitations of existing approaches and propose
further improvements. Experimental investigations are difficult to
make and involve approximations, for example very often the contact
parameters can only be estimated by indirect measurements of thermal
or electric resistances of compressed rough joints \citep{exp4,exp3}
or are mostly limited to measurements of real contact area under
special conditions \citep{exp1,exp2}. Therefore, numerical methods
are essential to acquire as much information as possible about the
contact problem at hand and infer general conclusions.

In spite of its effectiveness and versatility, the finite element
method (FEM) has been mainly applied in mechanics to solve contact
problems between rough surfaces in which the constitutive behavior
of the bulk is not linear elastic. For instance, the study of
elasto-plastic contact problems with roughness \citep{molinari},
where an explicit approach was used to reduce the high computational
cost, and the study involving frictional dissipative phenomena in
visco-elastic materials, where the energy dissipation in the bulk is
essential and can be well predicted by FEM \citep{wriggers}, are
worth mentioning.

In the linear elastic regime, when the multi-scale character of
roughness covering a wide spectrum of wavelengths is the main focus,
the use of the boundary element method (BEM) is historically
preferred over FEM \citep{A81,M94}. This is essentially due to the
fact that only the surface must be discretized and not the bulk.
Moreover, it is not necessary to adopt surface interpolation
techniques, like Bezier curves, to discretize the interface (see,
e.g., the rigorous studies in \cite[Ch. 9]{wriggers2} and
\citep{molinari}), which must be used with care to avoid smoothing
out artificially the fine scale geometrical features of roughness.

In the application of BEM, the frictionless contact problem between
two linear elastic rough surfaces is mathematically equivalent to
the problem of the normal contact between a rigid rough surface and
an elastic half-plane with equivalent elastic parameters, see
\citep{bounds} for a rigorous proof. The core of BEM is based on the
so-called Green's functions, that relate the displacement of a
generic point of the half-plane to the action of a concentrated
force on the surface caused by contact interactions. An integral
convolution of all the contact tractions provides the deformed
contact configuration. After introducing a discretization of the
half-plane consisting of a grid of boundary elements, the problem of
point-force singularity is solved numerically by using the
closed-form solution for a patch load acting on a finite-size
boundary element \citep[Ch. 3,4]{KJ}. The contact problem is then
set in terms of equalities and inequalities stemming from the
unilateral contact constraints and can be solved by constrained
optimization. In this regard, apart from the discretization error
intrinsic in any numerical method, BEM provides the highest
attainable accuracy for discrete problems \citep{PK99}. The basic
version of BEM can be also extended to solve rough contact problems
with friction \citep{li,pohrt} and between viscoelastic materials
\citep{putignano}.

With the aim of investigating the effect of roughness at multiple
scales, the availability of computational methods that can solve
large contact problems in an efficient and fast way is of crucial
importance. The size of the linear system of equations relating the
contact pressures to the normal deflections can be in fact quite
large, as it arises from high resolution profilometric surface
samples of 512$\times$512 heights and very large indentations.
Hence, the computational challenges regard two main aspects: $(i)$
efficiently solve the system of linear equations; $(ii)$ impose the
satisfaction of the unilateral contact constraints (contact
inequalities). Regarding the first issue, iterative methods like the
Conjugate Gradient algorithm or the Gauss-Seidel method
\citep{francis,BCC,BCC01} have been widely used. Alternatively, the
capabilities of multigrid or multilevel methods have been exploited
\citep{raous,PK99} to approximately solve the equation system on
coarse grids and then project the results on finer grids. Finally,
we mention the fast method and its variants based on the solution of
the linear system of equations in the Fourier space (see, e.g.,
\citep{kato,PK2000a,PK2000b,BHS,scaraggi,prodanov}).

Regarding the imposition of the contact inequalities,
\cite[p.149-150]{KJ} suggested to apply a greedy approach: after
solving the equation set for the unknown tractions, the boundary
elements for which these are negative (tensile) are excluded in a
following iteration from the assumed contact area and the
corresponding pressures set equal to zero. Johnson (1985)[p.149-150]
stated that ``experience confirms that repeated iterations converge
to a set of values of pressures which are positive where contact
takes place and zero otherwise". To the best of the authors'
knowledge, a rigorous proof of convergence of this method has not
been found in the literature. However, if valid, it allows to use
any numerical method to solve the unconstrained set of linear
equations and then impose a correction in a recursive way. Indeed,
this numerical approach has been successfully applied by many
authors, such as \cite{kubo} and \cite{BCC,BCC01} who used this
greedy approach in conjunction with a Gauss-Seidel iterative
algorithm for the solution of the unconstrained set of linear
equations, and \cite{karpenko} and \cite{BHS} who applied it
together with a numerical scheme based on the Fast Fourier Transform
(FFT).

In alternative to the greedy approach, \cite{PK99} proposed a
constrained Conjugate Gradient method based on the theory in
\citep[Ch. 2,3]{Hestenes} to solve the linear system of equations
and rigorously impose the satisfaction of the contact constraints.
For the solution of the system of equations, a multi-grid solution
scheme was proposed in \citep{PK99} and then a FFT algorithm was
considered in \citep{PK2000a,PK2000b}.

In this paper, we first examine the validity of the greedy approach
based on a monotonic elimination of tensile points. We show that
this approach usually finds the exact solution but, as we prove by a
counter-example, it may fail. Then, we show that other optimization
algorithms such as Non-Negative Least Squares (NNLS) and the
Alternative Direction Method of Multipliers (ADMM) can be used in
alternative to the greedy approach, by exploiting the equivalence
between the contact problem and quadratic programming with
unilateral non-negativity constraints. Moreover, we propose warm
starting techniques for the optimization algorithms that are
especially useful in case of a solution of a sequence of increasing
or decreasing displacements.

This paper provides a comprehensive comparison of the computational
performance of the greedy approach (used in conjunction with
different unconstrained solvers like the Conjugate Gradient, the
Gauss-Seidel iterative scheme, or the MATLAB's \texttt{mldivide}
solver\footnote{According to documentation, \texttt{mldivide} solves
linear systems with symmetric positive definite matrices by
computing a Cholesky factorization, see
\url{http://www.mathworks.it/help/MATLAB/ref/mldivide.html}}), of
the original constrained CG method by \citep{PK99}, and of novel
optimization algorithms that are able to exploit warm starts for
solving convex quadratic programs subject to non-negativity
constraints. As a main conclusion, the proposed NNLS algorithm with
warm start based on accelerated gradient projections (GPs) is found
to be one order of magnitude faster than the algorithm by
\cite{PK99} and two orders of magnitude faster than the greedy
approach.

Finally, by exploiting the morphological features of the contact
domain of fractal surfaces, we propose in this paper a cascade
multi-resolution algorithm that can further reduce computation time
by at least a factor two with respect to the NNLS algorithm with
accelerated GPs.

\section{Mathematical formulation}

In the framework of BEM, the normal
displacements $u(\bx)$ at any point of the half-plane identified by the position vector $\bx$ are related to the contact pressures $p(\by)$ at other points as follows \citep{KJ,jrB}:
\begin{equation}
\label{eq:Hup}
u(\bx)=\int_S H(\bx,\by)p(\by)\mbox{d}\by,
\end{equation}
where $H(\bx,\by)$ represents the displacement at a point $\bx$ due
to a surface contact pressure $p$ acting at $\by$
and $S$ is the elastic half-plane. For homogeneous, isotropic, linear elastic materials, the influence coefficients are:
\begin{equation}\label{gf}
H(\bx,\by)=\dfrac{1-\nu^2}{\pi
E}\dfrac{1}{\parallel\bx-\by\parallel},
\end{equation}
where $E$ and $\nu$ denote, respectively, the composite Young's
modulus and Poisson's ratio of half-plane, and $\|\cdot\|$
the standard Euclidean norm. The total contact force $P$ is the
integral of the contact pressure field
\begin{equation}
P=\int_S p(\bx)\mbox{d}\bx.
\end{equation}

\begin{figure}
\centering
\includegraphics[width=.8\textwidth,angle=0]{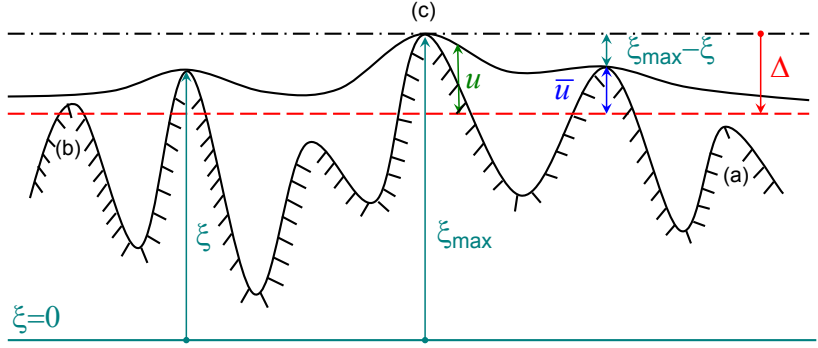}
\caption{Sketch of the contact problem between a rigid rough surface
and an elastic half-plane. Its deformed configuration corresponding
to the imposed far-field displacement $\Delta$ is depicted with a
black solid line. The red dashed line corresponding to a rigid-body
motion of the half-plane identifies the heights to
be included in the initial trial contact domain. Once Problem~\ref{prob:contact} is
solved we may have: ($i$) heights certainly not in contact from the
beginning, type $(a)$; ($ii$) heights loosing contact due to elastic
interactions, type (b); ($iii$) heights in contact, type
(c).}\label{fig:frame}
\end{figure}

By referring to Fig.~\ref{fig:frame}, in the following we define for
each surface point $\bx\in S$ its elevation $\xi(\bx)$, measured
with respect to a reference frame, and set $\xi_{\rm
max}\eqdef\max_{\bx\in S}\xi(\bx)$ the maximum elevation. The
indentation of the half plane at the points in contact is denoted by
$\bar{u}$, whereas a generic displacement along the surface is $u$.

We consider the following problem:
\begin{problem}
For a given far-field displacement $\Delta\geq0$ in the direction perpendicular to the undeformed half-plane, find the solution of the normal contact problem $u(\bx)$, $p(\bx)$ satisfying~\eqref{eq:Hup} and the unilateral contact (linear complementarity) conditions
\begin{subequations}
\label{unilateral}
\begin{align}
u(\bx)-\bar{u}(\bx,\Delta)&\ge 0,\\
p(\bx)&\ge 0,\\
(u(\bx)-\bar{u}(\bx,\Delta))p(\bx)&=0,
\end{align}
\end{subequations}
for all points $\bx\in S$, where contact tractions are positive when compressive.
\label{prob:contact}
\end{problem}
Introducing the quantity $w(\bx,\Delta)=u(\bx)-\bar{u}(\bx,\Delta)$,
Eq.\eqref{unilateral} can be rewritten as:
\begin{subequations}
\label{unilateral2}
\begin{align}
w(\bx,\Delta)&\ge 0,\\
p(\bx)&\ge 0,\\
w(\bx,\Delta)p(\bx)&=0.
\end{align}
\end{subequations}

Problem~\ref{prob:contact} is an infinite-dimensional linear
complementarity problem. We find a finite-dimensional approximate
solution by discretizing the surface as a square grid of spacing
$\delta$ consisting of $N\times N$ average heights. Let $S_{ij}$ be
the cell of area $\delta^2$ indexed by $i,j\in I_N$, with
$I_N\eqdef\{1,...,N\}\times\{1,...,N\}$. Let
$\bx_{i,j}\eqdef\int_{\bx\in S_{ij}}\bx \textrm{d}\bx$,
$\xi_{i,j}\eqdef\int_{\bx\in S_{ij}}\xi(\bx) \textrm{d}\bx$,
$p_{i,j}\eqdef\int_{\bx\in S_{ij}}p(\bx) \textrm{d}\bx$, and
$u_{i,j}\eqdef\int_{\bx\in S_{ij}}u(\bx) \textrm{d}\bx$ be,
respectively, the barycentric coordinate, average height, resultant
of the contact pressures, and the corresponding displacement on the
surface element $S_{ij}$. Consider the following discretized version
of~\eqref{eq:Hup}
\begin{equation}
u_{i,j}=\sum_{k=1}^{N}\sum_{l=1}^{N}H_{i-k,j-l}\,p_{k,l}
\label{eq:uij}
\end{equation}
for all $(i,j)\in I_N$, where the term $H_{i-k,j-l}$ is the Green
function used in~\eqref{eq:Hup} averaged over the elementary area
$\delta^2$, which corresponds to the displacement induced by a
uniformly loaded square:
\begin{equation}\label{green}
H_{i-k,j-l}=\dfrac{1}{\delta^2}
\int_{S_{ij}}\int_{S_{kl}}
H(\bx,\by)\text{d}\by\text{d}\bx,
\end{equation}
and
\begin{equation}
p_{k,l}\geq 0,\ \forall (k,l)\in I_N.
\label{eq:pkl}
\end{equation}
For instance, \cite{BCC} used the following approximation related to
a uniform pressure acting on a rounded patch of radius $\delta/2$:
\begin{equation}\label{green2}
H_{i-k,j-l}=\left\{
               \begin{array}{ll}
                 \dfrac{2}{E\pi\delta}, & \hbox{if $i=k$ and $j=l$} \\
                 \dfrac{2}{E\pi\delta}\arcsin\dfrac{\delta}{2\|\bx_{i,j}-\bx_{k,l}\|}, & \hbox{if $i\neq k$, $j\neq l$}
               \end{array}
             \right.
\end{equation}
but other formulae for a square patch can also be taken as in
\citep{pohrt}.

Let $\bar I_C\eqdef\{(i,j)\in I_N:\ \xi_{i,j}< \xi_{\rm
max}-\Delta\}$ be the set of indices corresponding to elements
$S_{ij}$ that are certainly not in contact (cf.
Fig.~\ref{fig:frame}), and hence
\begin{equation}
p_{k,l}=0, \forall (k,l)\in \bar I_C,
\label{eq:pinIC}
\end{equation}
let $m=\# \bar I_C$ be the number of elements of $\bar I_C$ and
$n=\# I_C$ the number of elements belonging to the initial trial
contact domain, $I_C\eqdef I_N\setminus \bar I_C$. The set $I_C$ is
only a superset of the set $I_C^*$ of actual contact points, since
the corrections to the displacements induced by elastic interactions
may induce lack of contact in some elements $(i,j)$, i.e.,
$u_{i,j}>\bar u_{i,j}$, where $\bar u_{i,j}\eqdef\Delta-\xi_{\rm
max}+\xi_{i,j}$ is the value of the compenetration of the height
corresponding to the element $(i,j)$ in the half-plane (see
Fig.~\ref{fig:frame}).

For a generic $(i,j)\in I_C$ corresponding to an element of the
surface which is potentially in contact with the elastic half-plane,
we denote by
\begin{equation}
w_{i,j}\eqdef u_{i,j}-\bar u_{i,j}\geq 0
\label{eq:wij}
\end{equation}
the corresponding elastic correction to the displacement. Clearly, it must hold that
\begin{equation}
w_{i,j}p_{i,j}=0,\ \forall (i,j)\in I_C
\label{eq:wp=0}
\end{equation}
since $w_{i,j}>0$ implies no contact between the surfaces and
therefore no pressure, while $p_{i,j}>0$ implies contact,
$u_{i,j}=\bar u_{i,j}$, or equivalently $w_{i,j}=0$.

By taking into account that $p_{k,l}=0$ for all $(k,l)\in \bar I_C$, Eq.~\eqref{eq:uij} can be recast as the following condition
\begin{equation}
w_{i,j}+\bar u_{i,j}=\sum_{(k,l)\in I_C} H_{i-k,j-l}\,p_{k,l},\ \forall (i,j)\in I_C,
\label{eq:wijH}
\end{equation}
which is limited to the nodes belonging to the initial trial contact
domain $I_C$, whose number of elements is in general significantly
smaller than those of $I_N$. The
relations~\eqref{eq:pkl}-\eqref{eq:wijH} can be recast in matrix
form as the following Linear Complementarity Problem
(LCP)~\citep{Cot92}:
\begin{subequations}
\begin{align}
&\bw=\bQ\bp-\bbu\\
&\bw\geq \bz,\ \bp\geq \bz,\ \bw'\bp=0,
\end{align}
\label{eq:LCP}%
\end{subequations}
where $\bw\in\rr^{n}$ is the vector of unknown elastic corrections
$w_{i,j}$, $(i,j)\in \bar I_C$, $\bw'$ denotes its transpose,
$\bp\in\rr^n$ is the vector of unknown average contact forces
$p_{i,j}$, $(i,j)\in I_C$, $\bbu\in\rr^n$ is the vector of
compenetrations $\bar u_{i,j}$, $(i,j)\in I_C$, and $\bQ=\bQ'$ is
the matrix obtained by collecting the compliance coefficients
$H_{i-k,j-l}$, for $(i,j),(k,l)\in I_C$. Due to the properties of
linear elasticity \citep[p.144]{KJ} we have that
\begin{equation}
\bQ=\bQ'\succ 0,
\label{eq:Q>0}
\end{equation}
that is $\bQ$ is a symmetric positive definite matrix (with the
additional property deriving from~\eqref{green2} of having all its
entries positive). After solving~\eqref{eq:LCP}, the vector
$\bu\in\rr^{n}$ of normal displacements $u_{i,j}$, $(i,j)\in I_C$,
is then simply retrieved as $\bu=\bbu+\bw$.

By the positive definiteness property~\eqref{eq:Q>0} of $\bQ$, we inherit
immediately the following important property~\citep[Th. 3.3.7]{Cot92}:
\begin{property}
\label{prop:uniqueness} The discretized version~\eqref{eq:pkl},
\eqref{eq:pinIC}-\eqref{eq:wijH} of Problem~\ref{prob:contact}
admits a unique solution $\bp$, $\bu$, for all $\Delta\geq0$.
\end{property}

The LCP problem~\eqref{eq:LCP} corresponds to the Karush-Kuhn-Tucker
(KKT) conditions for optimality of the following convex quadratic
program (QP)
\begin{subequations}\label{eq:QP}
\begin{align}
\text{min}_p&\; \dfrac{1}{2}\bp'\bQ\bp-\bbu'\bp\\
\text{s.t.}\; &\bp\ge\mathbf{0}
\end{align}
\end{subequations}
in that the solution $\bp$ of~\eqref{eq:QP} and its corresponding
optimal dual solution $\bw$ solve~\eqref{eq:LCP}, and vice versa.

Problem~\eqref{eq:QP} is consistent with former pioneering considerations by
\cite{kalker} and also summarized in \citep[p.151--152]{KJ}. In
fact, the contact pressures solving the unilateral contact problem
can be obtained by minimizing the total complementary energy $W$ of
the linear elastic system, subject to the constraint $p(\bx)\ge 0$, $\forall \bx\in\mathcal{S}$. For a
continuous system, the total complementary energy is
\begin{equation}
W=U-\int_{\mathcal{S}}p(\bx)\bar{u}(\bx,\Delta)\,\mbox{d}\bx,
\end{equation}
where $U$ is the internal complementary energy of the deformed
half-plane in contact. For linear elastic materials, we have:
\begin{equation}
U=\dfrac{1}{2}\int_{\mathcal{S}}p(\bx)u(\bx)\,\mbox{d}\bx.
\end{equation}
Although such an energy-based approach can be used to derive FEM formulations, it is
also possible to remain within BEM and introduce a surface
discretization as before. By invoking the averaged Green's functions
in~\eqref{green}, the discretized version of $\widetilde{W}$
leads to
\begin{equation}\label{objectf}
\widetilde{W}=\dfrac{1}{2}\sum_{(i,j)\in I_C}\sum_{(k,l)\in I_C}H_{i-k,j-l}\,p_{k,l}p_{i,j}-\sum_{(i,j)\in I_C}
p_{i,j}\bar u_{i,j}
\end{equation}
which represents a quadratic function of $\bp$ to be minimized,
under the constraints $p_{i,j}\geq0$, $\forall (i,j)\in I_C$, as in
\eqref{eq:QP}. Since it is unlikely that the contact area is known a
priori, the active set of nodes in contact results only after
solving problem~\eqref{eq:LCP} or equivalently~\eqref{eq:QP}.

A large variety of solvers for LCP and QP problems were developed in
the last 60 years~\citep{Bea55,Fle71,GI83,Cot92,SB94,PB14}, and is
still an active area of research in the optimization and control
communities. Historically, in the mechanics community, \cite{kalker}
proposed the simplex method, although it was found to be practical
only for relatively small $N$. More recent contributions adopt
algorithms to solve the unconstrained linear system of equations and
then correct the solution by eliminating the boundary elements
bearing tensile tractions \citep{francis,BCC,BCC01}, or use a
constrained version of the Conjugate Gradient (CG) algorithm
\citep{PK99}. These methods are simply initialized by considering
arbitrary nonnegative entries in $\bp$, without taking advantage of
the monotonic increase (or decrease) of pressures by increasing (or
decreasing) the far-field displacement, an important property
guaranteed by rigorous elasticity theorems \citep{barber74}. The
history of pressures is saved during a contact simulation and it is
easy to access and use and it can be beneficial to save computation
time.

Next section presents effective optimization algorithms for solving
the QP problem~\eqref{eq:QP} and compares their performance with
respect to the Greedy CG method. Contrary to the latter, not only
the considered QP have the guaranteed property of always converging
to the unique solution $\bp$, $\bu$ for any given $\Delta\geq 0$,
but also the history of loading can be more efficiently taken into
account as a warm-start, with a significant saving of computation
time.

\section{Optimization algorithms}
Since now on, we use the subscript $i$ to denote the $i$-th
component of a vector or the $i$-th row of a matrix, the subscript
$\IC$ to denote the subvector obtained by collecting all the components $i\in\IC$ of a vector (or all the rows $i$ of a matrix), and the double subscript $\IC,\IC_1$ to
denote the submatrix obtained by collecting the $i$-th row and
$j$-th column, for all $i\in\IC$, $j\in\IC_1$.

\subsection{Greedy methods}

A greedy method corresponds to solve problem~\eqref{eq:QP} by
iteratively solving the unconstrained linear system of equations
$\bw=\bQ\bp-\bar\bu=\bz$ with respect to $\bp$ and increasingly
zeroing negative elements of $\bp$ until the condition $\bp\geq \bz$
is satisfied. By construction we obtain $\bw'\bp=0$. The method is
described in Algorithm~\ref{algo:greedy}, in which a standard
Conjugate Gradient employed to solve the unconstrained linear system
of equations. Steps 2.1-2.4 can be replaced by any other algorithm
for solving the linear system of equations, like the Gauss-Seidel
iterative scheme as in \citep{BCC,BCC01}, the MATLAB's
\texttt{mldivide} solver, or even the FFT algorithm as in
\citep{karpenko,BHS}.

\begin{algorithm}
\caption{Greedy method with Conjugate Gradient (greedy CG)}
\label{algo:greedy} \vspace*{.1cm}\hrule\vspace*{.1cm}
~~\textbf{Input}: Matrix $\bQ=\bQ'\succ 0$, vector $\bbu$; initial
guess $\bp$ and initial active set $\IC\subseteq\{1,\ldots,n\}$ such
that $\bp_{\{1,\dots,n\}\setminus\IC}=\bz$; maximum number $K_{\rm
max}$ of iterations, tolerance $\epsilon>0$.
\vspace*{.1cm}\hrule\vspace*{.1cm}
\begin{enumerate}
\item $i\leftarrow 0$; $\bar{\IC}\leftarrow
\{1,\dots,n\}\setminus\IC$;
\item \textbf{while} ($i\leq K_{\rm max}$ \textbf{and} $\min(\bp)<-\epsilon$)
\textbf{or} $i=0$ \textbf{do}:
\begin{enumerate}
\item [(2.1)]$\bw_{\IC}\leftarrow\bQ_{\IC,\IC}\bp_{\IC}-\bbu_{\IC}$;
\item [(2.2)]$n_w\leftarrow\|\bw_{\IC}\|_2$;
\item [(2.3)]$\bb_{\IC}\leftarrow-\bw_{\IC}$
\item [(2.4)]\textbf{while} $n_w>\epsilon$ \textbf{and} $i\leq K_{\rm max}$ \textbf{do}:
\begin{enumerate}
\item [(2.4.1)]$\bs_{\IC}\leftarrow\bQ_{\IC,\IC}\bb_{\IC}$;
\item [(2.4.2)]$\bp_{\IC}\leftarrow \bp_{\IC}-\frac{\bw_{\IC}'\bb_{\IC}}{\bb_{\IC}'\bs_{\IC}}\bb_{\IC}$;
\item [(2.4.3)]$\bbw_{\IC}\leftarrow \bQ_{\IC,\IC}\bp_{\IC}-\bbu_{\IC}$;
\item [(2.4.4)]$\bb_{\IC}\leftarrow -\bbw_{\IC}+\frac{\bbw_{\IC}'\bs_{\IC}}{\bb_{\IC}'\bs_{\IC}}\bb_{\IC}$;
\item [(2.4.5)]$\bw_{\IC}\leftarrow\bbw_{\IC}$;
\item [(2.4.6)]$n_w\leftarrow \|\bw_{\IC}\|_2$;
\item [(2.4.7)]$i\leftarrow i+1$;
\end{enumerate}
\item [(2.5)]\textbf{for} $j\in\IC$ \textbf{do}:
\begin{enumerate}
\item [(2.5.1)]\textbf{if} $\bp_j<-\epsilon$ \textbf{then} $\bp_j\leftarrow 0$;
$\IC\leftarrow\IC\setminus\{j\}$; $\bar{\IC}\leftarrow
\bar{\IC}\cup\{j\}$;
\end{enumerate}
\end{enumerate}
\item $\bp^*\leftarrow \bp$;
\item $\bu^*_\IC=\bbu_\IC$, $\bu_{\bar\IC}^*\leftarrow
\bQ_{\bar\IC,\IC}\bp_\IC$;
\item \textbf{end}.
\end{enumerate}
\vspace*{.1cm}\hrule\vspace*{.1cm} ~~\textbf{Output}: Contact force
vector $\bp^*$ and normal
displacement vector $\bu^*$. 
\vspace*{.1cm}\hrule\vspace*{.1cm}
\end{algorithm}

Assuming that the prescribed initial $\bp$ and $\IC$ are such that
$p_j=0$ for all $j\in\{1,\ldots,n\}\setminus \IC$, and $K_{\rm max}$
is sufficiently large, the output of the greedy algorithm leads to a
contact pressure vector $\bp^*$ and a normal displacement vector
$\bu^*$ satisfying $\bu^*=\bQ\bp^*$, $\bp^*\geq \bz$,
$(\bu^*-\bbu)'\bp^*=0$. In fact, condition $\bp^*\geq \bz$ is
guaranteed by the condition in Step~2 up to $\epsilon$ precision. By
letting $\bw^*\eqdef\bu^*-\bbu$, at termination of the algorithm we
have $\bw^*_\IC=\bQ_{\IC,\IC}\bp^*_{\IC}-\bbu_{\IC}=\bz$ because of
the solution of the CG method (Step 2.4), or equivalently
$\bu^*_\IC= \bbu_\IC$ (cf. Step 5). By setting $\bu^*_{\bar
\IC}\eqdef\bQ_{\bar\IC,\IC}\bp_\IC$ in Step 5, and recalling that
$\bp^*_{\bar\IC}=\bz$, we have
\[
    \matrice{c}{\bw^*_\IC\\\bw^*_{\bar \IC}} 
    =\matrice{cc}{\bz & \bz\\\bQ_{\bar\IC,\IC}&\bz} 
    \matrice{c}{\bp^*_\IC\\\bz 
    }+\matrice{c}{\bz\\-\bbu_{\bar \IC}}=
    \matrice{cc}{\bQ_{\IC,\IC}&\bQ_{\IC,\bar\IC}\\\bQ_{\bar\IC,\IC}&\bQ_{\bar\IC,\bar\IC}}\matrice{c}{\bp^*_\IC\\\bp^*_{\bar \IC}}+\matrice{c}{-\bbu_{\IC}\\-\bbu_{\bar \IC}}
\]
and hence $\bu^*=\bw^*+\bbu=\bQ\bp^*$. The complementarity condition
$(\bu^*-\bbu)'\bp^*=(\bw^*)'\bp^*=0$ follows by construction, as Step 2.4 zeroes
all the components of $\bw^*_j$ that correspond to nonnegative $\bp^*_j$,  $\forall j\in\IC$, and zeroes all the components $\bp^*_j$ that correspond to possible nonzero components $\bw^*_j$, $\forall j\in\bar\IC$.

However, to the best of the authors' knowledge, no formal proof
exists that the condition $\bw^*_{\bar\IC}\geq \bz$ is satisfied
after the algorithm terminates, i.e., that $\bu^* \geq \bbu$. If the
algorithm is applied to randomly generated $\bbu$ vectors and $\bQ$
positive definite matrices with positive coefficients, in many cases
the LCP is not solved exactly. In contact mechanics, the only
evidence that this condition is satisfied has been shown in
simulations (see, e.g.,~\citep{BHS}). Indeed, we obtained the
following counterexample in which the greedy method failed in
getting the solution also for $\bQ$ whose coefficients are given by
Eq.\eqref{green2}\footnote{The MATLAB routine of the counterexample
is available for download at
\url{http://musam.imtlucca.it/counterexample.m}}.

\begin{figure}
\centering \subfigure[Random
$\bbu$ as input]{\includegraphics[width=.48\textwidth,angle=0]{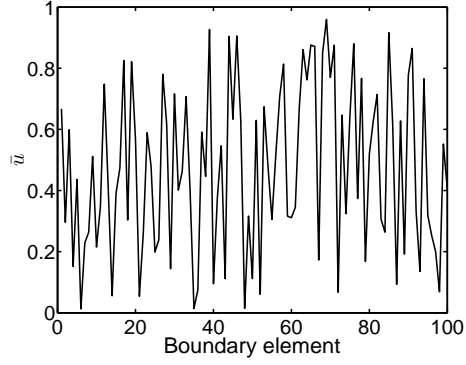}\label{fig2a}}\\
\subfigure[Computed $\bw^*_{\bar\IC}$]{\includegraphics[width=.48\textwidth,angle=0]{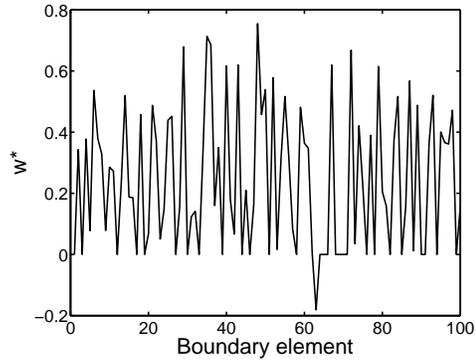}\label{fig2b}}\\
\subfigure[Computed
$\bp/E$]{\includegraphics[width=.48\textwidth,angle=0]{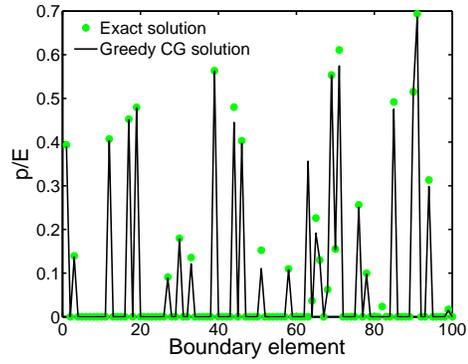}\label{fig2c}}
\caption{Counterexample showing that the greedy CG method fails in
getting the correct solution ($\delta=1$ a.u. of $L$, as $\bu$ and
$\bw^*$; $E=0.01$ $F/L^2$). Green dots correspond to the correct
contact forces satisfying the LCP and are obtained by using the NNLS
method, Sec. 3.2}\label{fig2}
\end{figure}

\begin{example}
{\normalfont Consider a square mesh with grid spacing $\delta$
consisting of $N\times N$ boundary elements indexed by $(i,j)\in
I_N$, $I_N=\{1,\dots,N\}\times \{1,\dots,N\}$. Suppose that all the
boundary elements are included in the initial trial contact domain
$I_C$ $(n=N\times N)$ by assigning a positive value of $\bbu_{i,j}$
to all elements. This may represent a situation where a cluster of
densely packed heights comes into contact. Since $\bbu_{i,j}$
depends on the height field $\xi_{i,j}$, which is a random variable,
for the sake of generality we extract the values $\bbu_{i,j}$
randomly from a uniform distribution in the interval $(0,1)$. The
matrix $\bQ$ is assembled according to~\eqref{green2}. By running a
sequence of 100 random simulations, we find that in approximately
$40\%$ of the simulations the greedy method provides a solution
which violates the condition $w^*_{i,j}\geq 0$ in at least one
element. This lack of getting the right solution is observed for any
size $n$ of the problem. One of the wrong results obtained for
$n=100$ is shown in Fig.~\ref{fig2}. The assigned random values of
$\bbu$ are plotted in Fig.~\ref{fig2a} for the sequence of boundary
elements (from 1 to 100) composing the mesh. The solution $\bw^*$
presents a negative entry in one single element (element 62 in
Fig.~\ref{fig2b}). The computed contact forces are compared in
Fig.~\ref{fig2c} with the values corresponding to the exact solution
of the problem (green dots) obtained by using the NNLS algorithm
presented in Section~\ref{sec:NNLS}, that is proven to satisfy the
LCP conditions~\eqref{eq:LCP} exactly. Although just one value of
$\bw^*$ is negative, the overall solution is affected by this
violation. We observe in fact a false contact detection for the
element number 62 violating the condition $\bw^*_{i,j}>0$, a contact
not detected (element 81) and 7 contact forces significantly
underestimated with respect to the exact ones. }\hfill$\blacksquare$
\label{ex:greedy-bad}
\end{example}

For less densely packed boundary elements belonging to $I_C$, for
instance with a minimum distance of $2\delta$ between them instead
of $\delta$ as in Example~\ref{ex:greedy-bad}, the algorithm was
found to always provide a solution satisfying the condition
$\bw^*\geq\bz$. Other benchmark tests considering a deterministic
smooth variation of $\bbu$, as in case of an indentation by a smooth
sphere or by a flat punch, did not show any convergence problem to
the solution as well, although the boundary elements in contact are
densely packed as in the counterexample shown before. In conclusion,
although it is likely that the diagonally dominant property of the
matrix $\bQ$ plays a role in the robustness of the algorithm, it
remains an open problem to find exact mathematical requirements for
$\bQ$ and $\bbu$ that guarantee the greedy method to provide a
solution satisfying $\bw^*\geq \bz$, so that all the LCP
conditions~\eqref{eq:LCP} are met.

Therefore, as a word of caution, the reliability of the greedy
method should be carefully checked in case of applications of BEM to
contact problems governed by other forms of $\bQ$, as in the case of
contact with an anisotropic or an inhomogeneous half-space, or in
the presence of multiple fields.

Another drawback of the algorithm is the difficulty to warm start
the method with a proper choice of the initial active set $\IC$.
Since at Step 2.5.1 the number of elements in the sequence $\IC$ is
decreased by removing negative enough components $\bp_j$ of the
current solution vector, i.e., eliminating the points bearing
tensile (negative) forces, in a monotonic way (no index $j$ that has
been removed from $\IC$ can be added back), a safe cold start is to
set $\IC=\{1,\ldots,n\}$ and pick up a vector $\bp\geq\textbf{0}$,
usually a vector with arbitrary non-negative numbers. The history of
contact forces obtained during the solution of a sequence of imposed
displacements is not taken into account by the method to accelerate
its convergence, although we know that contact forces are
monotonically increasing functions of the far-field displacement. In
any case, for a complex sequence of loading with an increased or
decreased far-field displacement, any warm starting on forces cannot
be implemented in the method, since the elimination of contact
points is irreversible.

\subsection{Constrained Conjugate Gradient}

A constrained CG algorithm was proposed by \cite{PK99} based on the
theory by \cite[Ch. 2,3]{Hestenes} to solve the linear system of
equations and rigorously impose the satisfaction of the contact
constraints. Algorithm \ref{algo:conCG} has been applied by
\cite{PK99} to simulations under load control. However, it can be
used also for displacement control. The condition for convergence
set by \cite{PK99} in terms of relative variation in the local
contact forces from an iteration to the next has been recast in
terms of the error in the local contact displacements. The two
criteria are completely equivalent.

This constrained CG algorithm does not remove the points bearing
tensile forces from the active set. Therefore, the size of the
linear system of equations is not reduced during the iterations,
increasing the computation time for its solution. On the other hand,
the method assures the satisfaction of the LCP conditions
\eqref{eq:LCP} and it is found to convergence with a reduced number
of iterations as compared to the Greedy CG algorithm. Although not
investigated in \citep{PK99}, it can be warm started in case of a
sequence of loading steps by considering both an initial trial
contact domain and a set of contact pressures derived from the
previous converged solution. The FFT method can be used to
accelerate step (2.7) as in \citep{PK2000a}.

\begin{algorithm}
\caption{Constrained Conjugate Gradient} \label{algo:conCG}
\vspace*{.1cm}\hrule\vspace*{.1cm} ~~\textbf{Input}: Matrix
$\bQ=\bQ'\succ 0$, vector $\bbu$, initial guess $\bp\ge\bz$, initial
active set $\IC=\{1,\ldots,n\}$; maximum number $K_{\rm max}$ of
iterations, tolerance $\epsilon>0$.
\vspace*{.1cm}\hrule\vspace*{.1cm}
\begin{enumerate}
\item $i\leftarrow 0$, $n_{w,\rm{old}}=1$, $d=0$, $err=+\infty$;
\item $\bw\leftarrow\bQ\bp-\bbu$;
\item \textbf{while} ($i\leq K_{\rm max}$ \textbf{and} $err>\epsilon$):
\begin{enumerate}
\item [(3.1)]\textbf{if} $i=0$ \;\textbf{then} $\bt\leftarrow\bw$
\;\textbf{else}: $\bt\leftarrow
\bw+d\dfrac{n_w}{n_{w,\rm{old}}}\bt_{\rm{old}}$;
\item [(3.2)] $\tau=\dfrac{\bw'\bt}{\bt'\bQ\bt}$;
\item [(3.3)] $\bp\leftarrow\bp-\tau\bt$;
\item [(3.4)] $\forall j\in\IC: p_j\leftarrow \max\{p_j,0\}$;
\item [(3.5)] Find $I_{ol}=\{j\in \IC: p_j=0,w_j<0\}$;\\
\textbf{if} $I_{ol}=\emptyset$ \textbf{then} $d=1$ \textbf{else}
$d=0$; $p_j\leftarrow p_j-\tau w_j$, $\forall j\in I_{ol}$;
\item [(3.6)] $\IC\leftarrow \{j:p_j>0\}\cup I_{ol}$;
\item [(3.7)] $\bt_{\rm{old}}\leftarrow \bt$, $n_{w,\rm{old}}\leftarrow n_w$;
\item [(3.8)] $\bw\leftarrow\bQ\bp-\bbu$;
\item [(3.9)] $n_w=\|\bw\|_2$;
\item [(3.10)] $err\leftarrow |n_w-n_{w,\rm{old}}|/n_{w,\rm{old}}$;
\item [(3.11)]$i\leftarrow i+1$;
\end{enumerate}
\item $\bp^*\leftarrow \bp$; $\bu^*=\bQ\bp^*$;
\item \textbf{end}.
\end{enumerate}
\vspace*{.1cm}\hrule\vspace*{.1cm} ~~\textbf{Output}: Contact force
vector $\bp^*$ and normal displacement vector $\bu^*$.
\vspace*{.1cm}\hrule\vspace*{.1cm}
\end{algorithm}

\subsection{Non-Negative Least Squares (NNLS)}
\label{sec:NNLS}

In this section we show how a QP problem with positive definite
Hessian matrix having the special form \eqref{eq:QP} can be
effectively solved as a nonnegative least squares problem.

Thanks to property~\eqref{eq:Q>0}, matrix $\bQ$ admits a Cholesky
factorization $\bQ=\bC'\bC$. Hence we can \emph{theoretically}
recast problem~\eqref{eq:QP} as the Non-Negative Least Squares
(NNLS) problem:
\begin{subequations}\label{eq:NNLS}
\begin{align}
\text{min}_p&\; \dfrac{1}{2}\|\bC\bp-\bC^{-T}\bbu\|_2^2\\
\text{s.t.}\; &\bp\ge\mathbf{0}
\end{align}
\end{subequations}
A simple and effective active-set method for solving the NNLS
problem~\eqref{eq:NNLS} is the one in~\cite[p.161]{LH74}, that is
extended here in Algorithm~\ref{algo:NNLS} to directly solve
\eqref{eq:QP} without explicitly computing the Cholesky factor $\bC$
and its inverse $\bC^{-1}$ and to handle warm starts. After a finite
number of steps, Algorithm~\ref{algo:NNLS} converges to the optimal
contact force vector $\bp^*$ and returns the normal displacement
vector $\bu^*$ whose components $p_{i,j}$, $u_{i,j}$ satisfy
$p_{i,j}\geq 0$, $u_{i,j}\geq \bar u_{i,j}$, $(u_{i,j}-\bar
u_{i,j})p_{i,j}=0$, and~\eqref{eq:wijH}, $\forall (i,j)\in I_C$.

The method is easy to warm start in case of a loading scenario
consisting of an alternating sequence of increasing or decreasing
far-field displacements. The contact forces determined for a given
imposed displacement are used to initialize vector $\bp$. Due to the
monotonicity of the contact solution, this initialization is
certainly much closer to the optimal solution $\bp^*$ than a zero
vector. This usually significantly reduces the iterations of the
method to convergence. Such a warm start has a fast implementation
requiring a projection of the forces of the points belonging to
$I_C^*(\Delta_k)$ to the same points of the trial domain
$I_C^*(\Delta_{k+1})$ for a new imposed far field displacement
$\Delta_{k+1}$. For an increasing far-field displacement, i.e.,
$\Delta_{k+1}>\Delta_{k}$ the forces in the elements belonging to
$I_C^*(\Delta_{k+1})-I_C^*(\Delta_k)$ are simply initialized equal
to zero. In the numerical experiments of
Section~\ref{sec:simulations}, the time required for this projection
will be added to the global solution time for a consistent
comparison with the greedy method with cold start and with the
constrained CG algorithm.

\begin{algorithm}
\caption{Non-Negative Least Squares (NNLS)} \label{algo:NNLS}
\vspace*{.1cm}\hrule\vspace*{.1cm} ~~\textbf{Input}: Matrix
$\bQ=\bQ'\succ 0$, vector $\bbu$, initial guess $\bp$; maximum
number $K_{\max}$ of iterations, tolerance $\epsilon>0$.
\vspace*{.1cm}\hrule\vspace*{.1cm}
\begin{enumerate}
\item $\IC\leftarrow\{i\in\{1,\ldots,n\}:\ \bp_i>\bz\}$;
$init\leftarrow\textsc{FALSE}$; $k\leftarrow 0$;
\item \textbf{if} $\IC=\emptyset$ \textbf{then} $init\leftarrow\textsc{TRUE}$;
\item $\bw\leftarrow \bQ\bp-\bbu$;
\item \textbf{if} (($\bw\geq -\epsilon$ \textbf{or} $\IC=\{1,\ldots,n\}$) \textbf{and}
$init=\textsc{TRUE}$) \textbf{or} $k\ge K_{\max}$ \textbf{then go
to} Step 13;
\item \textbf{if} $init=\textsc{TRUE}$ \textbf{then} $i\leftarrow\arg\min_{i\in\{1,\ldots,n\}\setminus\IC} \bw_i$; $\IC\leftarrow\IC\cup\{i\}$;\\ \textbf{else} $init\leftarrow\textsc{TRUE}$;
\item $\bs_I \leftarrow$ solution of the linear system $\bQ_I\bs_I=\bbu_I$
\item \textbf{if} $\bs_\IC\geq -\epsilon$ \textbf{then} $\bp\leftarrow \bs$ and \textbf{go to} Step 3;
\item $j\leftarrow\arg\min_{h\in\IC:\ \bs_h\leq \bz}\left\{\frac{\bp_h}{\bp_h-\bs_h}\right\}$;
\item $\bp\leftarrow \bp+\frac{\bp_j}{\bp_j-\bs_j}(\bs-\bp)$;
\item $\II\leftarrow\{h\in\IC: \bp_h=\bz\}$;
\item $\IC\leftarrow\IC\setminus\II$; $k\leftarrow k+1$;
\item \textbf{go to} Step 6;
\item $\bp^*\leftarrow \bp$;
\item $\bu^*\leftarrow \bw+\bbu$;
\item \textbf{end}.
\end{enumerate}
\vspace*{.1cm}\hrule\vspace*{.1cm} ~~\textbf{Output}: Contact force
vector $\bp^*$ and normal displacement vector $\bu^*$ satisfying
$\bu^*=\bQ\bp$, $\bu^*\geq \bbu$, $\bp^*\geq \bz$,
$(\bu^*-\bbu)'\bp=0$. \vspace*{.1cm}\hrule\vspace*{.1cm}
\end{algorithm}

Note that Step 6 of Algorithm~\ref{algo:NNLS} is equivalent to Step
2.4 of Algorithm~\ref{algo:greedy} and has been performed by using
the MATLAB's \texttt{mldivide} solver. This step can be accelerated
by the use of an approach based on the FFT (for its implementation,
see e.g. \citep{BHS}). Alternatively, since the set $\II$ changes
incrementally during the iterations of the algorithm, more efficient
iterative QR~\cite[Chap. 24]{LH74} or LDL$^T$~\cite{Bem14b}
factorization methods can be employed.

\subsubsection{Warm-started NNLS via accelerated Gradient Projection (NNLS+GP)}
An alternative method to solve Problem~\eqref{eq:QP} is to use an
accelerated gradient projection (GP) method for
QP~\citep{Nes83,PB14}. Because of the simple nonnegative constraints
in~\eqref{eq:QP}, rather than going to the dual QP formulation as
in~\citep{PB14}, we formulate the GP problem directly for the primal
QP problem~\eqref{eq:QP}. Numerical experiments have shown slow
convergence of a pure accelerated GP method to solve~\eqref{eq:QP}.
However, we can use the method to \emph{warm start}
Algorithm~\ref{algo:NNLS}, as described in Algorithm~\ref{algo:GP}.
If Algorithm~\ref{algo:GP} is executed $(K>0)$, it returns a vector
$\bp$ that is immediately used as an input to
Algorithm~\ref{algo:NNLS}, otherwise one can simply set $\bp=\bz$
(cold start). As shown in Section~\ref{sec:simulations}, GP
iterations provide large benefits in warm starting the NNLS solver,
therefore allowing taking the best advantages of the two methods:
quickly getting in the neighborhood of the optimal solution (GP
iterations of Algorithm~\ref{algo:GP}) and getting solutions up to
machine precision after a finite number of iterations (the
active-set NNLS Algorithm~\ref{algo:NNLS}).

\begin{algorithm}
\caption{Accelerated Gradient Projection (GP)}
\label{algo:GP}
\vspace*{.1cm}\hrule\vspace*{.1cm}
~~\textbf{Input}: Matrix $\bQ=\bQ'\succ 0$ and its Frobenius norm $L$, vector $\bbu$, initial guess $\bp$, number $K$ of iterations.
\vspace*{.1cm}\hrule\vspace*{.1cm}
\begin{enumerate}
\item $\bar{\bp}\leftarrow\bp$;
\item \textbf{for} $i=0,\ldots,K-1$ \textbf{do}:
\begin{enumerate}
\item [(2.1)]$\beta=\max\{\frac{i-1}{i+2},0\}$;
\item [(2.2)]$\bs=\bp+\beta(\bp-\bar{\bp})$;
\item [(2.3)]$\bw=\bQ\bs-\bbu$;
\item [(2.4)]$\bar{\bp}\leftarrow\bp$;
\item [(2.5)]$\bp\leftarrow\max\{\bs-\frac{1}{L}\bw,\bz\}$;
\end{enumerate}
\item \textbf{end}.
\end{enumerate}
\vspace*{.1cm}\hrule\vspace*{.1cm} ~~\textbf{Output}: Warm start for
contact force vector $\bp$ and elastic correction vector $\bw$.
\vspace*{.1cm}\hrule\vspace*{.1cm}
\end{algorithm}

\subsection{Alternating Direction Method of Multipliers (ADMM)}
The QP problem~\eqref{eq:QP} can also be solved by the Alternating
Direction Method of Multipliers (ADMM), which belongs to the class
of augmented Lagrangian methods. The reader is referred to
~\citep{BPCPE11} for mathematical details. The method treats the
QP~\eqref{eq:QP} as the following problem
\begin{equation}
\ba{rl}
    \min_{\bp,\bs}&\frac{1}{2}\bp'\bQ\bp-\bar\bu'\bp+g(\bs)\\
    \st&\bp=\bs
\ea
\label{eq:ADMM}
\end{equation}
where
\[
    g(\bs)=\left\{\ba{rcl}0&\mbox{if}&\bs\geq \bz\\+\infty&\mbox{if}&\bs<\bz\ea\right.
\]
Then, the augmented Lagrangian function
\[
    L_\rho(\bp,\bs,\bw)=\frac{1}{2}\bp'\bQ\bp-\bar\bu'\bp+g(\bs)+\bw'(\bp-\bs)+
    \frac{\rho}{2}\|\bp-\bs\|_2^2
\]
is considered, where $\rho>0$ is a parameter of the algorithm. The basic ADMM algorithm consists of the following iterations:
\begin{equation}
\ba{rcl}
    \bp^{k+1}&=&\arg\min_\bp L_\rho(\bp,\bs^k,\bw^k)\\
    \bs^{k+1}&=&\arg\min_\bs L_\rho(\bp^{k+1},\bs,\bw^k)\\
    \bw^{k+1}&=&\bw^k+\rho(\bp^{k+1}-\bs^{k+1})
\ea
\label{eq:ADMM-iters}
\end{equation}
A scaled form with over-relaxation of the ADMM iterations~\eqref{eq:ADMM-iters}
is summarized in Algorithm~\ref{algo:ADMM}. The algorithm is guaranteed
to converge asymptotically to the solution $\bp^*$, $\bu^*$ of the problem.
The over-relaxation parameter
$\alpha>1$ is introduced to improve convergence, typical values for $\alpha$
suggested in~\citep{BPCPE11} are $\alpha\in[1.5,1.8]$.

A warm start of the algorithm that takes into account the loading
history is possible in a way analogous to that described for the
NNLS approach of Section~\ref{sec:NNLS}. However, as an additional complexity, also an initialization for the dual variable vector $\bw$ must be provided,
possibly obtained by projecting the solution obtained for a certain
$\Delta_{k}$ to that for $\Delta_{k+1}$.

\begin{algorithm}
\caption{Alternative Direction Method of Multipliers (ADMM)}
\label{algo:ADMM} \vspace*{.1cm}\hrule\vspace*{.1cm}
~~\textbf{Input}: Matrix $\bQ=\bQ'\succ 0$, vector $\bbu$, initial
guesses $\bp$, $\bw$, parameter $\rho>0$, over-relaxation parameter
$\alpha>1$, maximum number $K_{\rm max}$ of iterations, tolerance
$\epsilon>0$. \vspace*{.1cm}\hrule\vspace*{.1cm}
\begin{enumerate}
\item $\bM\leftarrow (\frac{1}{\rho}\bQ+\bI)^{-1}$;
\item $\bw_\rho\leftarrow -\frac{1}{\rho}\bw$;
\item $\bs\leftarrow \bp$;
\item $i\leftarrow 0$;
\item \textbf{while} ($i\leq K_{\rm max}$ \textbf{and} $\|\bp-\bs\|_\infty>\epsilon$)
\textbf{or} $i=0$ \textbf{do}:
\begin{enumerate}
\item [(5.1)]$\bs\leftarrow \bM(\bp-\bw_\rho-\frac{1}{\rho}\bbu)$;
\item [(5.2)]$\bar{\bs}\leftarrow\alpha \bs+(1-\alpha) \bp$;
\item [(5.3)]$\bp\leftarrow\max\{\bar{\bs}+\bw_\rho,\bz\}$;
\item [(5.4)]$\bw_\rho\leftarrow\bw_\rho+\bar{\bs}-\bp$;
\item [(5.5)]$i\leftarrow i+1$;
\end{enumerate}
\item $\bp^*\leftarrow \bp$;
\item $\bu^*\leftarrow \bbu-\rho\bw_\rho$;
\item \textbf{end}.
\end{enumerate}
\vspace*{.1cm}\hrule\vspace*{.1cm} ~~\textbf{Output}: Contact force
vector $\bp^*$ and normal displacement vector $\bu^*$ satisfying
$\bu^*=\bQ\bp$, $\bu^*\geq \bbu$, $\bp^*\geq \bz$,
$(\bu^*-\bbu)'\bp=0$. \vspace*{.1cm}\hrule\vspace*{.1cm}
\end{algorithm}

\section{Performance comparison of the algorithms}
\label{sec:simulations}

The optimization algorithms presented in the previous section are
herein applied to the frictionless normal contact problem between a
numerically generated pre-fractal rough surface and a half-plane, in
order to compare their performance in terms of number of iterations
required to achieve convergence and computation time.

\begin{figure}
\centering
\includegraphics[width=.8\textwidth,angle=0]{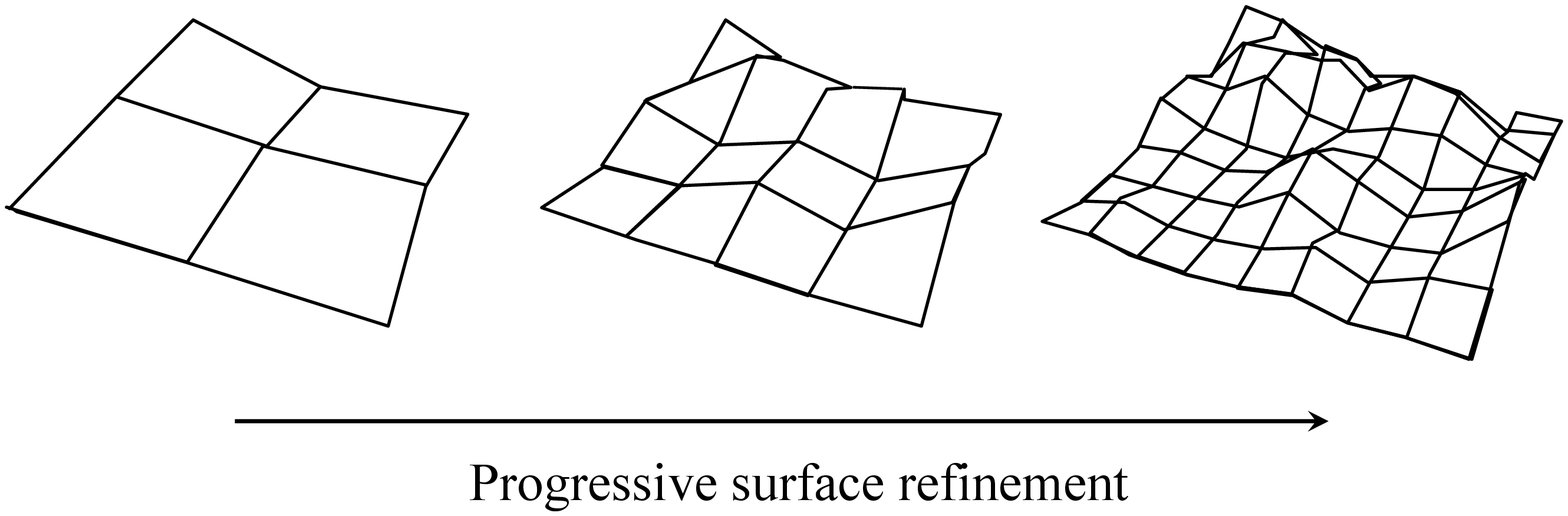}
\caption{Rough surfaces with multi-scale roughness and different
resolution, numerically generated by the random midpoint displacement
algorithm.}\label{fig:rmd}
\end{figure}

The random midpoint displacement algorithm \citep{PS88} is used to
generate the synthetic height field of surfaces with multiscale
fractal roughness, i.e., with a power spectral density (PSD)
function of the height field of power-law type. The surface with a
given resolution (pre-fractal) is realized by a successive
refinement of an initial coarse representation by adding a sequence
of intermediate heights whose elevation is extracted from a Gaussian
distribution with a suitable rescaled variance, see a qualitative
sketch in Fig.~\ref{fig:rmd}. Several applications of the method to
model rough surfaces for contact mechanics simulations are available
in \citep{ZBP04,ZBP07,PC10}.

In particular, we consider a test problem consisting of a surface
with Hurst exponent $H=0.7$, lateral size $L=100$ $\mu$m and 512
heights per side, which corresponds to the highest discretization
used to sample real surfaces with a confocal profilometer, like the
Leica DCM3D available at the Multi-scale Analysis of Materials
(MUSAM) Laboratory of IMT Lucca, Italy. Similar discretizations are
obtained in case of AFM. The surface is brought into contact with an
elastic half-plane under displacement control. Ten displacement
steps are imposed to reach a maximum far-field displacement which is
set equal to $(\xi_{\max}-\xi_{\text{ave}})/2$, where $\xi_{\max}$
and $\xi_{\text{ave}}$ are the maximum and the average elevations of
the rough surface, respectively. All the simulations are carried out
with the server 653745-421 Proliant DL585R07 from Hewlett Packard
with 128 GB Ram, 4 processors AMD Opteron 6282 SE 2.60 GHz with 16
cores running MATLAB R2014b.

\begin{figure}
\centering
\includegraphics[width=.6\textwidth,angle=0]{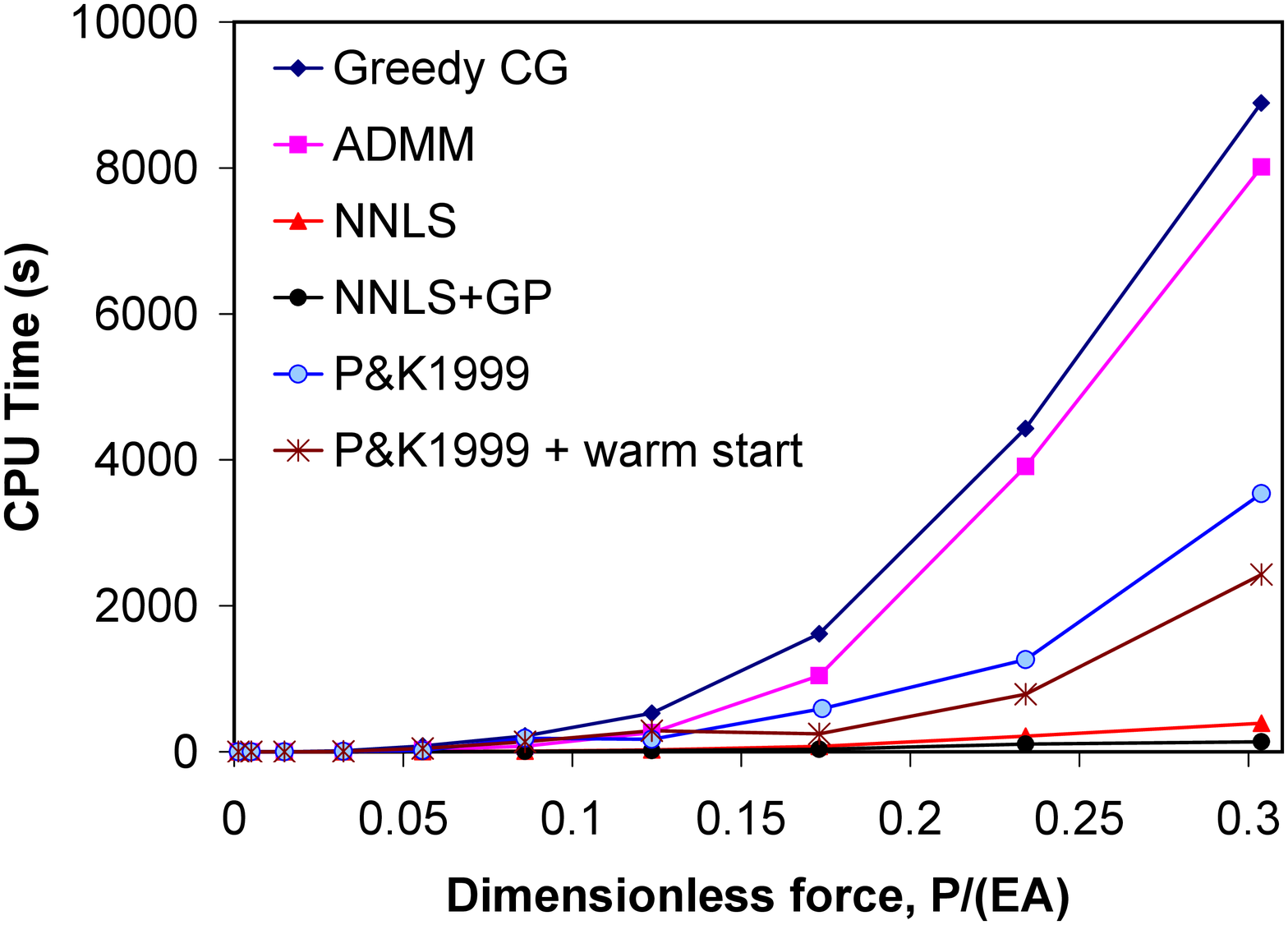}
\caption{Comparison between the optimization algorithms in terms of
computation time.}\label{fig4}
\end{figure}

The parameters for the Greedy CG method are the maximum number of
iterations $K_{\max}=1\times 10^{5}$ and the convergence tolerance
$\epsilon=1\times 10^{-8}$. The contact forces are initialized at
zero (cold start). The constrained CG method also considers
$K_{\max}=1\times 10^{5}$ and the same tolerance $\epsilon=1\times
10^{-8}$. Both the original version by \cite{PK99} (labeled P\&K1999
in Fig.~\ref{fig4}) and its warm-started variant (labeled P\&K1999 +
warm start in Fig.~\ref{fig4}) are considered.

For the NNLS algorithm (Algorithm 3) we adopt the warm start
strategy based on the projection of contact forces from the solution
corresponding to a previous displacement step. Alternatively, for
NNLS+GP, 100 gradient projections are used to initialize vector
$\bp$. For the ADMM method we use $\alpha=1.5$, $\rho=1$,
$K_{\max}=3\times 10^{3}$ and $\epsilon=10^{-8}$. The total number
$n$ of optimization variables is varying with $\Delta$ and therefore
with the force level. For the highest indentation we have $n=35555$.
Warm starting the algorithm is achieved by projecting primal
variables as for the NNLS and dual variables $\bw$ as well. The
projection simply consists of assigning the values of $p_{i,j}^*$
and $w^*$ of the boundary elements in contact for the step
$\Delta_k$ to the same boundary elements belonging to the trial
contact domain $I_C$ corresponding to the higher indentation
$\Delta_{k+1}$.

Once convergence is achieved for each imposed far-field
displacement, the optimization algorithms provide the same normal
force $P$ and contact domains, with small roundoff errors due to
finite machine precision. The CPU time
required by each method to achieve convergence are shown in
Fig.~\ref{fig4} vs. the dimensionless normal force $P/(EA)$, where
$E$ is the Young's modulus and $A=L^2$ is the nominal contact area.
The best performance is achieved by the application of the NNLS
method with 100 gradient projections (GP), which is 26 times faster than the original constrained CG method by \cite{PK99} and
about two orders of magnitude faster than
the ADMM and the Greedy CG algorithms.

As outlined in the introduction, the Greedy method can be used in
conjunction with other algorithms for solving the unconstrained
linear system of equations (Step 2.4) than the CG algorithm.
Although an extensive comparison of different solvers of linear
systems of equations with positive definite matrices is outside the
scope of this paper, we tested the Greedy algorithm after replacing
the CG Step 2.4 with the optimized built-in \texttt{mldivide}
function of MATLAB, or with the Gauss-Seidel algorithm, as proposed
in \cite{BCC,BCC01}.

\begin{figure}
\centering {\includegraphics[width=.6\textwidth,angle=0]{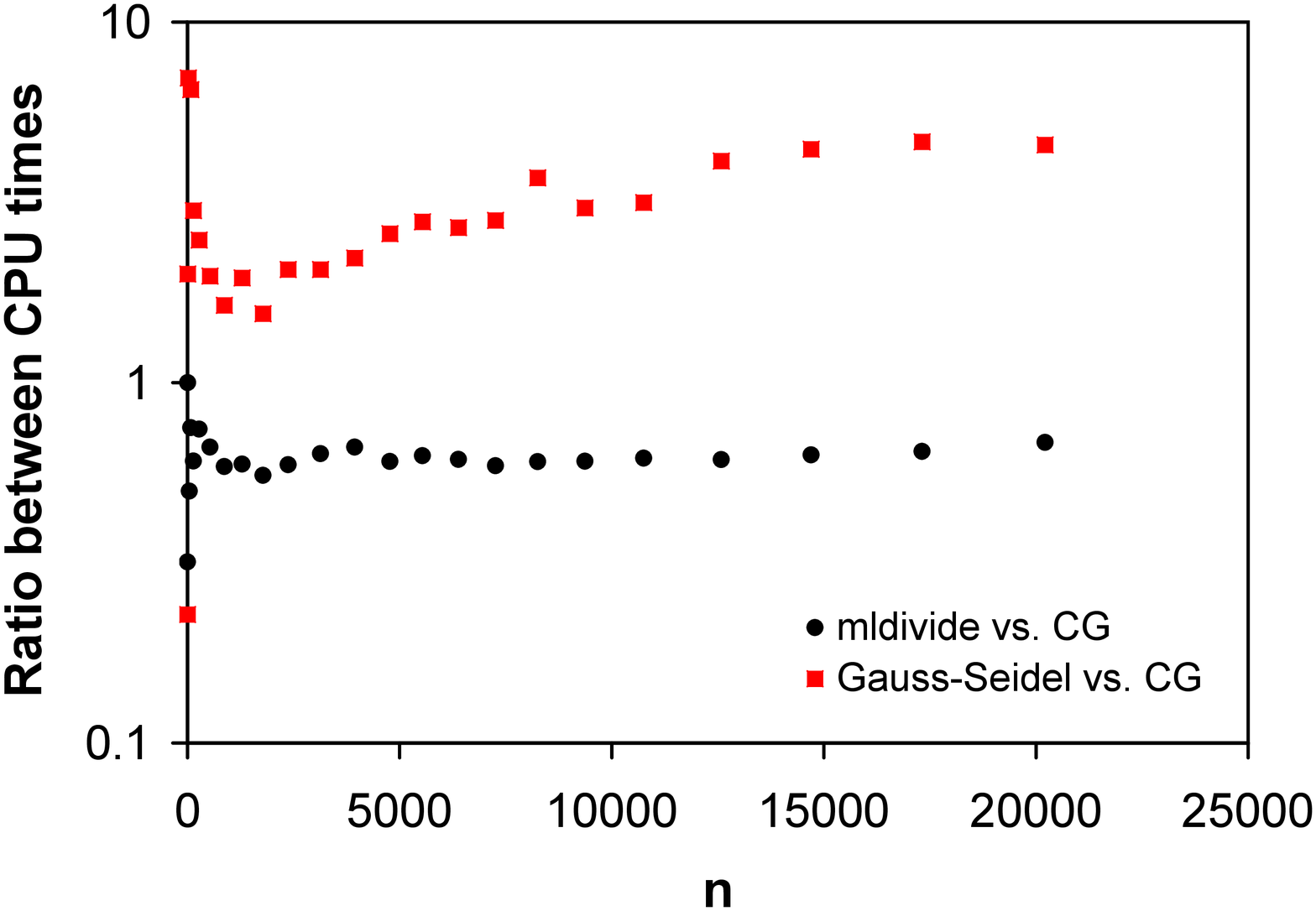}}
\caption{Computation times for the Greedy method for different sizes
$n$ of the contact superset $I_C$: CG vs. MATLAB's \texttt{mldivide}
solver and CG vs. Gauss-Seidel algorithm.}\label{fig5}
\end{figure}

The MATLAB's \texttt{mldivide} solver (which employs the Cholesky
factorization) leads to a reduction of computation time of
$30-40\%$, almost regardless of the size of the system $n$, see
Fig.~\ref{fig5}. Even with this gain in computation speed, the
overall performance is still quite far from that of the NNLS
Algorithm~\ref{algo:NNLS} on the platform used for the tests.
Moreover, the MATLAB solver leads to an error of lack of memory for
$n>20000$, a serious problem for large systems that is not suffered
by the CG solver described in Step 2.4 of
Algorithm~\ref{algo:greedy}. The Gauss-Seidel algorithm does not
suffer for the lack of memory but it is about 3 times slower than
the CG method.

\begin{figure}
\centering {\includegraphics[width=.6\textwidth,angle=0]{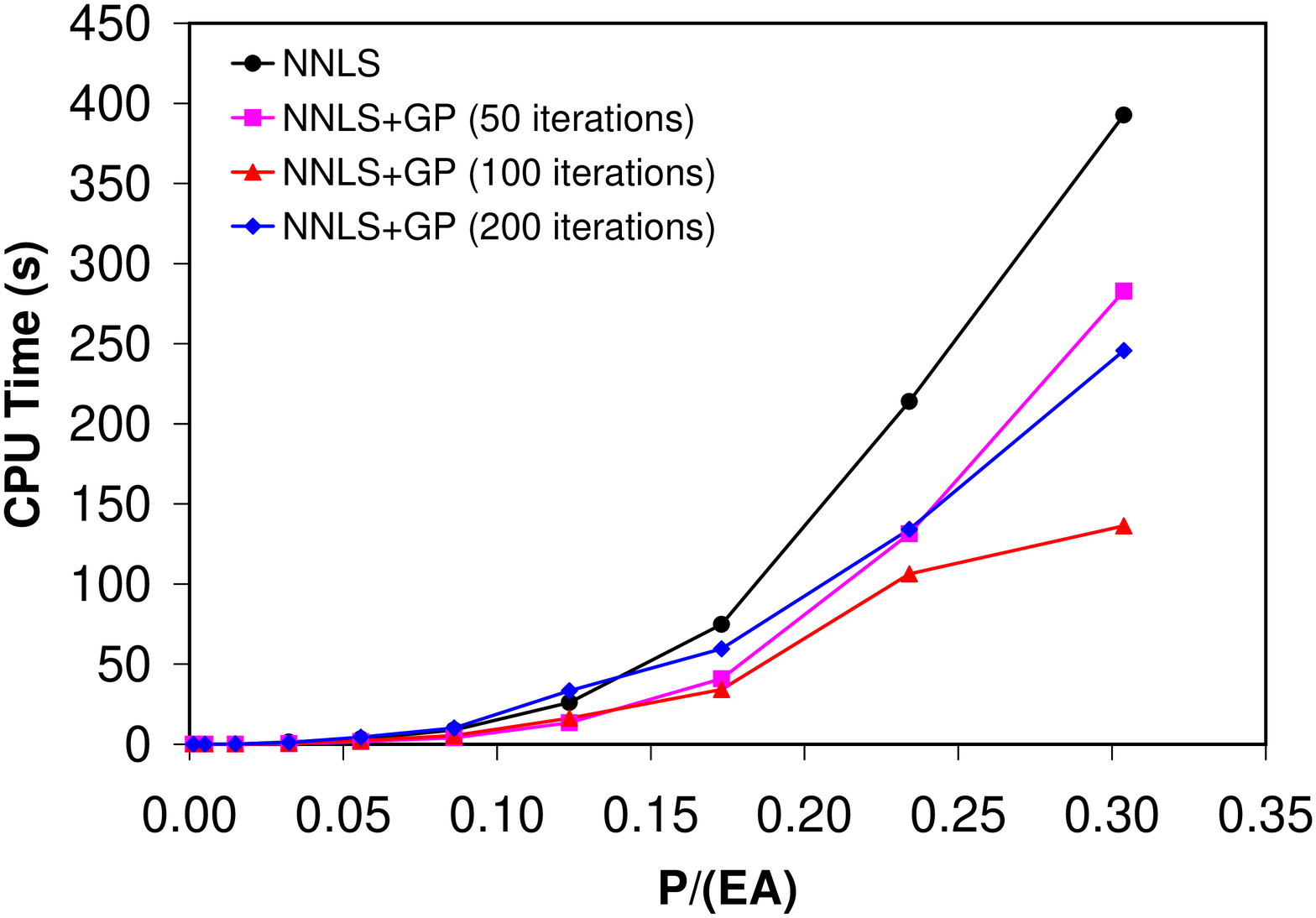}}
\caption{Computation times of the NNLS algorithm depending on the
number $K$ of gradient projection (GP) iterations.}\label{fig6}
\end{figure}

The effect of the number $K$ of GP iterations applied before the NNLS algorithm is investigated in Fig.~\ref{fig6} for the same test problem whose results were shown in Fig.~\ref{fig4}. By increasing $K$
from 0 to 100 we observe a reduction in the total computation time
due to a decrease in the number of iterations requested by the NNLS
algorithm to achieve convergence thanks to a better initial guess of
$\bp$. However, a further increase in $K$ (see, e.g., the blue curve
in Fig.~\ref{fig6} corresponding to $K=200$ iterations) does not
correspond to further savings of CPU time. This is due to the fact
that the number of NNLS iterations was already reduced to its
minimum for $K=100$ GP iterations, so that the application of
further gradient projections are just leading to additional CPU time
without further benefit.

\section{Cascade multi-resolution (CMR) method}

\subsection{Algorithm}

A further speed-up of computation time, as compared to the NNLS
method, can be achieved by improving the criterion for the guess of
the initial set $I_{C}$ of points in contact. The standard criterion
based on checking the interpenetration of the surface heights into
the half-plane in case of a rigid body motion is the most
conservative. However, at convergence, only a small subset $I_C^*$
of that initial set is actually in contact. Therefore, a better
choice of the initial trial contact domain would reduce the size of
the system of linear equations with an expected benefit in terms of
computation time.

\begin{figure}
\centering {\includegraphics[width=.75\textwidth,angle=0]{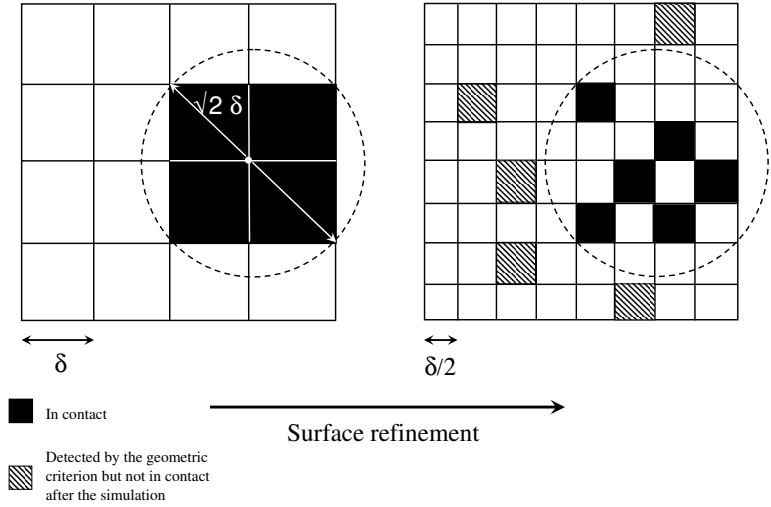}}
\caption{A sketch illustrating the property of lacunarity of the
contact domain: the real contact area progressively diminishes by
refining the surface, until vanishes in the fractal limit of
$\delta\to 0$. This implies that some boundary elements detected by
the rigid-body interpenetration criterion (dashed grey elements) can
be neglected a priori since they are outside the real contact area
corresponding to the coarse scale contact solution.}\label{fig7}
\end{figure}

As shown in \citep{BCC} via numerical simulations on pre-fractal
surfaces with Hurst exponent $H>0.5$ and different resolution, by
refining the surface height field via a recursive application of the
random midpoint displacement algorithm the real contact area of each
surface representation decreases by reducing $\delta$, as
illustrated in the sketch in Fig.~\ref{fig7}. In the fractal limit
of $\delta\to 0$, the real contact area vanishes. Therefore, this
property of lacunarity implies that the heights that are not in
contact for a coarser surface representation are not expected to
come into contact by a successive refining of the height field, for
the same imposed far-field displacement.

Therefore, as a better criterion, the initial trial contact domain
can be selected by retaining, among all the heights selected by the
rigid body interpenetration check, only those located within the
areas of influence of the nodes belonging to the contact domain of a
coarser representation of the rough surface for the same imposed
displacement $\Delta$.

As graphically shown in Fig.~\ref{fig7}, an area of influence of a
given node in contact can be defined by the radius $\sqrt{2}\delta$,
where $\delta$ is the grid size of the coarser surface
representation. Since the criterion is not exact, it is convenient
to consider a multiplicative factor $h$ larger than one for the
radius defining the nodal area of influence. It is remarkable to
note that this numerical scheme can be applied recursively to a
cascade of coarser representations of the same rough surface. As a
general trend, computation time is expected to drastically diminish
by increasing the number of cascade projections. However, the
propagation of errors due to the wrong exclusion of heights that
would actually make contact cannot be controlled by the algorithm
and it is expected to increase with the number of projections as
well. The advantage of the method is represented by the fact that,
in addition to saving computation time with respect to that required
by the NNLS algorithm to solve just the contact problem for the
finest surface, all the contact predictions for the coarser scale
representations of the same surface will be available for free,
which is a useful result for the multi-scale characterization of
contact problems. Moreover, the CMR method can be used in
conjunction with any of the optimization algorithms presented in the
previous sections.

The algorithm is illustrated in Algorithm \eqref{algo:CMR}.

\begin{algorithm}
\caption{Cascade multi-resolution (CMR) algorithm} \label{algo:CMR}
\vspace*{.1cm}\hrule\vspace*{.1cm} ~~\textbf{Input}: $s=1,\dots,l$
surface representations with different resolution or grid spacing
$\delta(s)$; area of influence parameter $h\ge 1$.
\vspace*{.1cm}\hrule\vspace*{.1cm}
\begin{enumerate}
\item \textbf{for} $s=1,\dots,l$ \textbf{do}:
\begin{enumerate}
\item [(1.1)]Determine $I_C(s)=\{(i,j)\in I_N(s):\xi_{i,j}\ge\xi_{\max}(s)-\Delta\}$;
\item [(1.2)]\textbf{if} $s=1$ \textbf{then}  $I_{C,p}(s)=I_C(s)$\\
\textbf{else} $I_{C,p}(s)=\{(i,j)\in I_C(s):r_{i-k,j-l}=\|\bx_{i,j}-\bx_{k,l}\|\le h\delta(s-1)\}$, $\forall (k,l)\in I_C^*(s-1)$\\
\textbf{end}
\end{enumerate}
\item Construct $\bQ$ based on the projected trial contact domain $I_{C,p}(s)$;
\item Apply optimization algorithms (e.g., NNLS) and determine $\bp^*$,
$\bu^*$, $I_C^*(s)$;
\item\textbf{end}.
\end{enumerate}
\vspace*{.1cm}\hrule\vspace*{.1cm}
\end{algorithm}

\subsection{Validation in case of numerically generated and real rough surfaces}

To assess the computational performance of the approach described in
Section 5.1, the CMD method is applied in conjunction with the NNLS
algorithm to pre-fractal surfaces with different $H$ numerically
generated by the RMD method. As an example, the lateral size is
$100$ $\mu$m for all the surfaces and the finest resolution whose
contact response has to be sought corresponds to 256 heights per
side. The method requires the storage of the coarser representations
of such surfaces that are in any case available by the RMD algorithm
during its various steps of random addition.

We apply the cascade of projections starting by a coarser
representation of the surfaces with only 16 heights per side and
then considering 32, 64, 128 and finally 256 heights per side. A
parameter $h=2$ has been used for the definition of the area of
influence. The solution of the contact problem for the surface with
16 heights per side is obtained in an exact form since it is the
starting point of the cascade, whereas the contact predictions for
the finer surface representations can be affected by an error
intrinsic in the criterion. The approximate predictions for the
surface with 256 heights per side are compared with the reference
solution corresponding to the application of the NNLS algorithm with
warm start directly to the finest representation of the rough
surface.

\begin{figure}
\centering
\subfigure[$H=0.3$]{\includegraphics[width=.48\textwidth,angle=0]{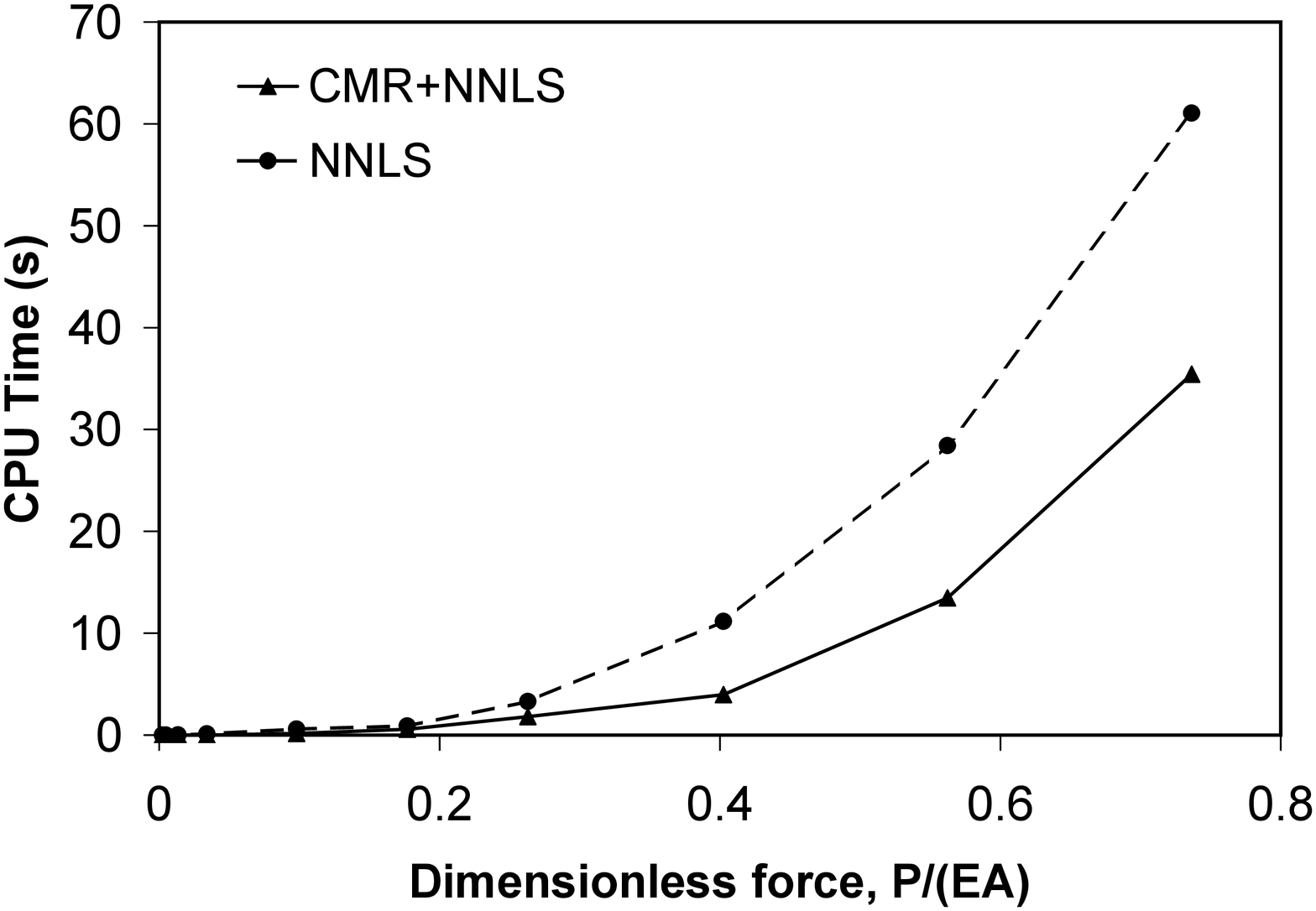}\label{fig8a}}\quad
\subfigure[$H=0.5$]{\includegraphics[width=.48\textwidth,angle=0]{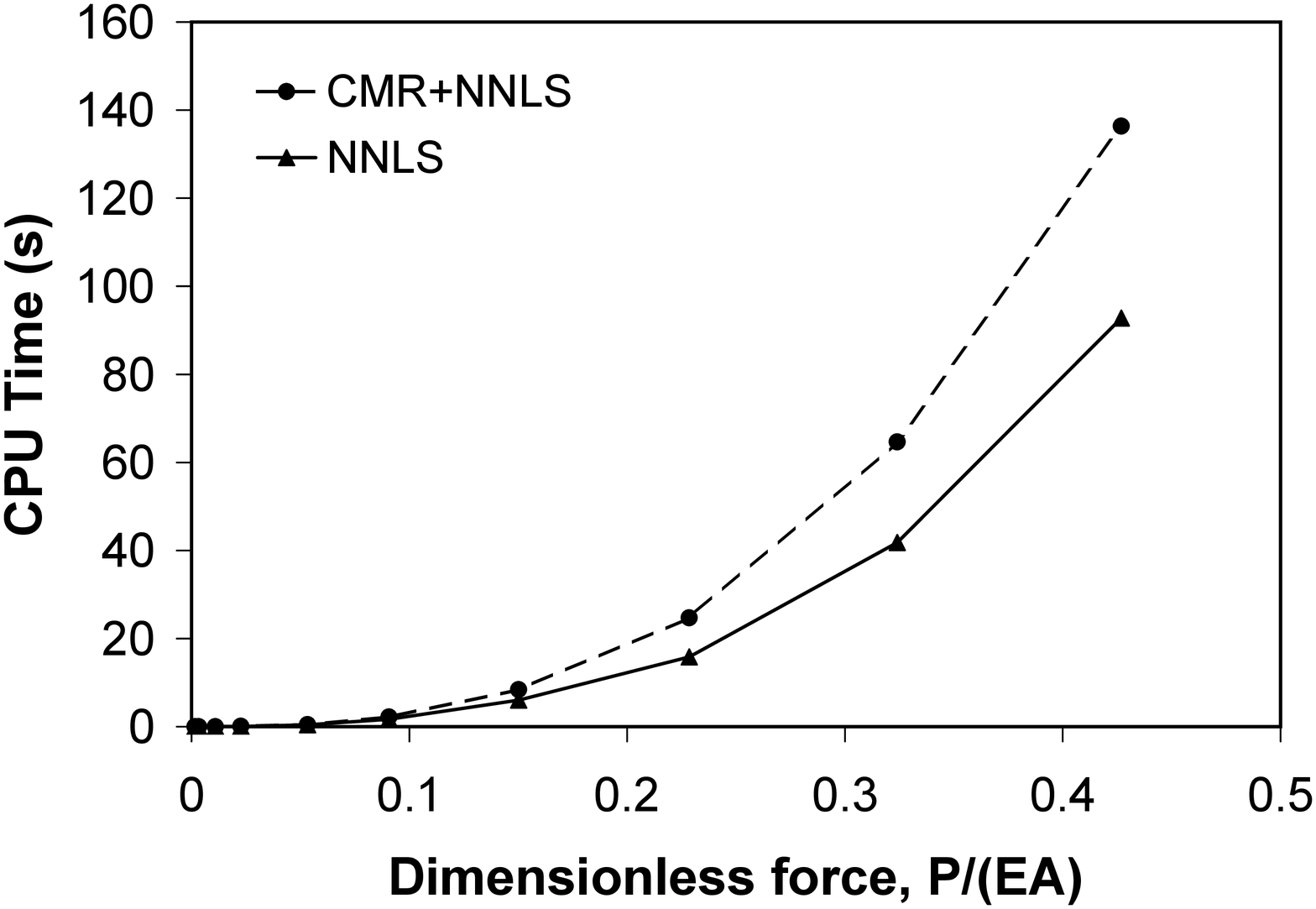}\label{fig8b}}\\
\subfigure[$H=0.7$]{\includegraphics[width=.48\textwidth,angle=0]{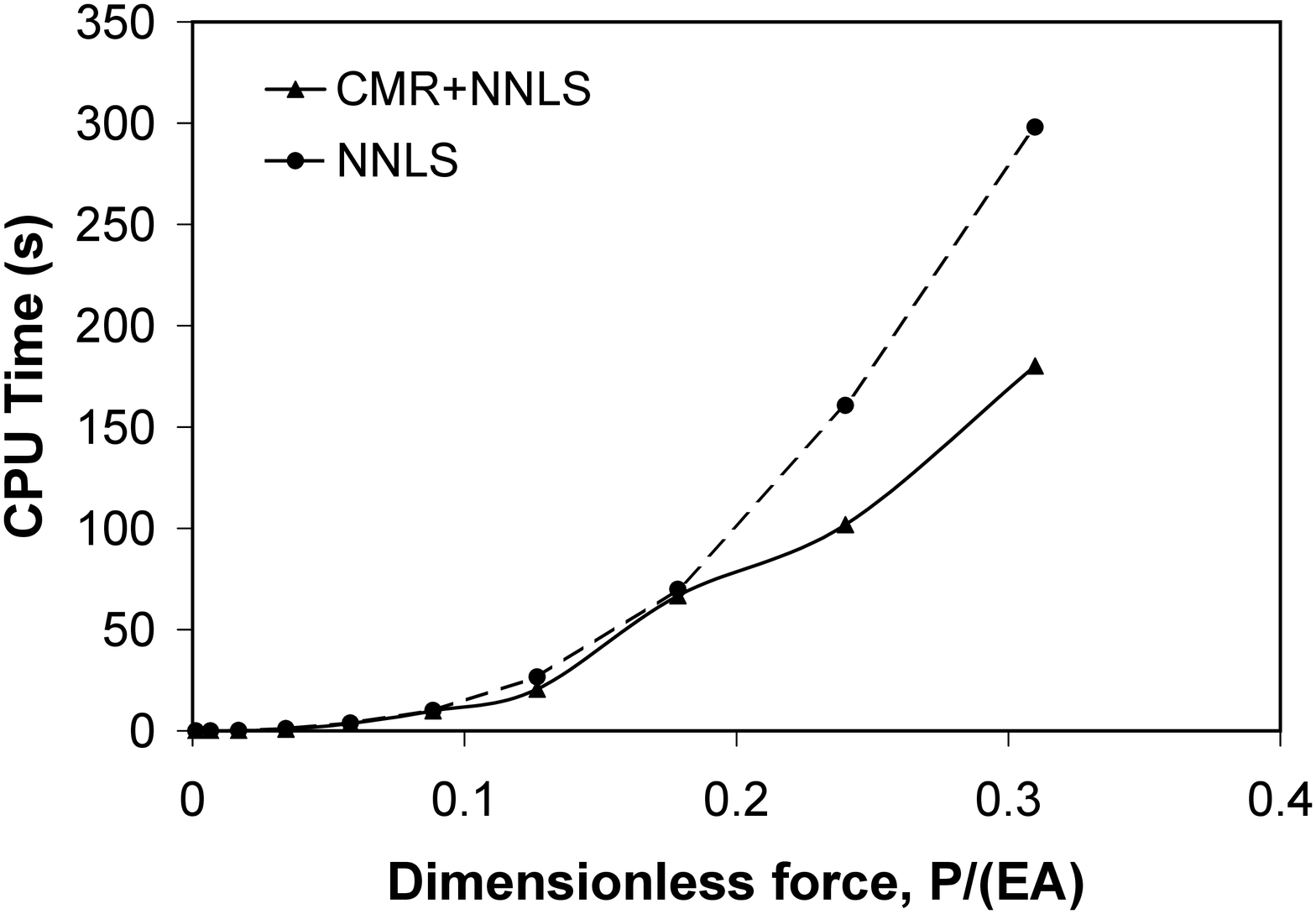}\label{fig8c}}\quad
\subfigure[Relative
error]{\includegraphics[width=.47\textwidth,angle=0]{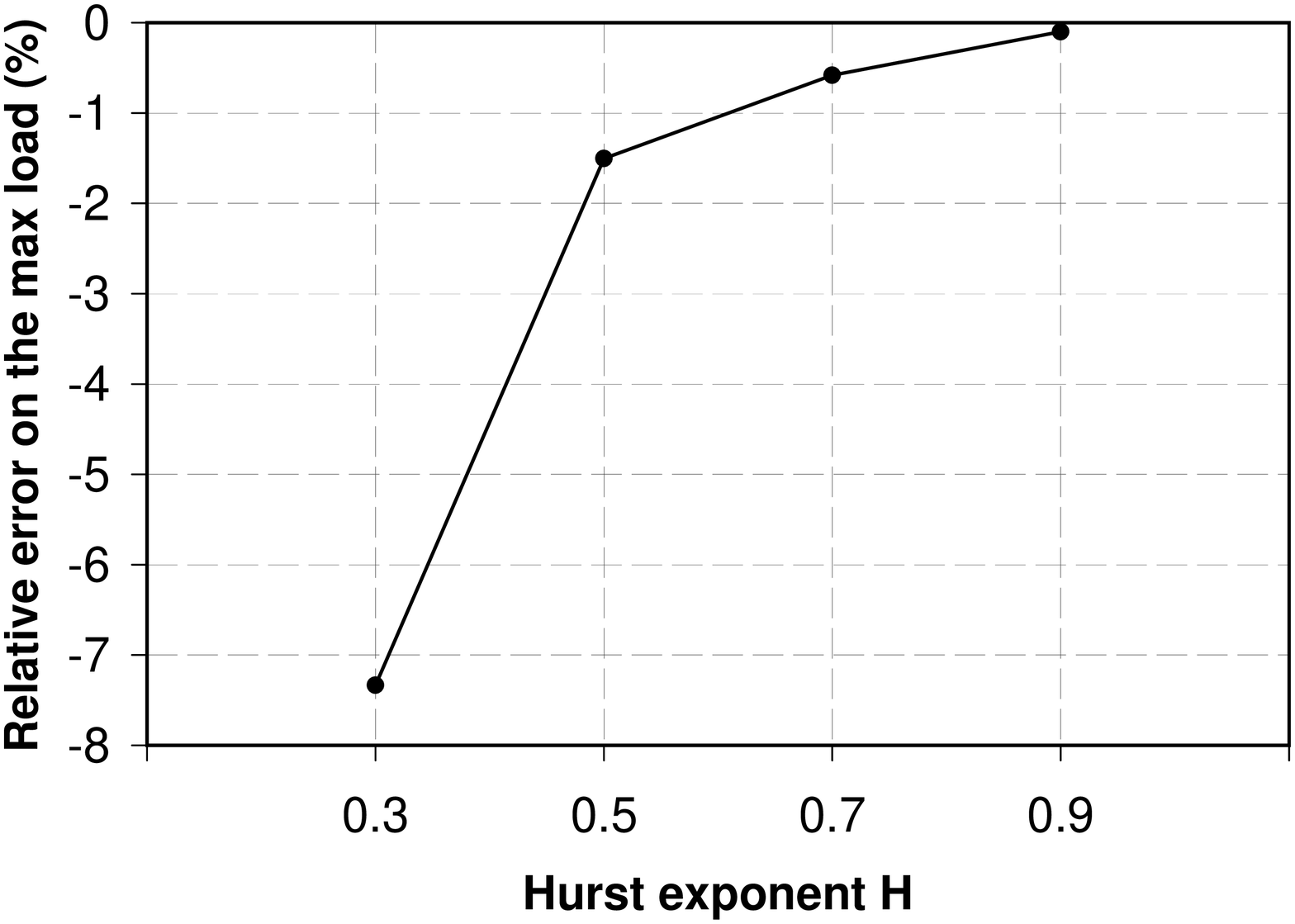}\label{fig8d}}
\caption{Performance of the CMR+NNLS method applied to numerically
generated fractal surfaces with different Hurst exponent $H$,
$h=2$.}\label{fig8}
\end{figure}

The computation time of the CMR+NNLS solution is the sum of the CPU
time required to solve all the coarser surface representations and
it is found to be much less than the CPU time required by the NNLS
algorithm to solve just one single surface with the finest
resolution, see Fig.~\ref{fig8}, where we observe a reduction of
$50\%$ in CPU time almost regardless of $H$. The relative error in
the computation of the maximum normal force between the predicted
solution and the reference one is a rapid decreasing function of
$H$, as shown in Fig.~\ref{fig8d}. Considering that real surfaces
have often a Hurst exponent $H>0.5$, this is very promising.

\begin{figure}
\centering {\includegraphics[width=.6\textwidth,angle=0]{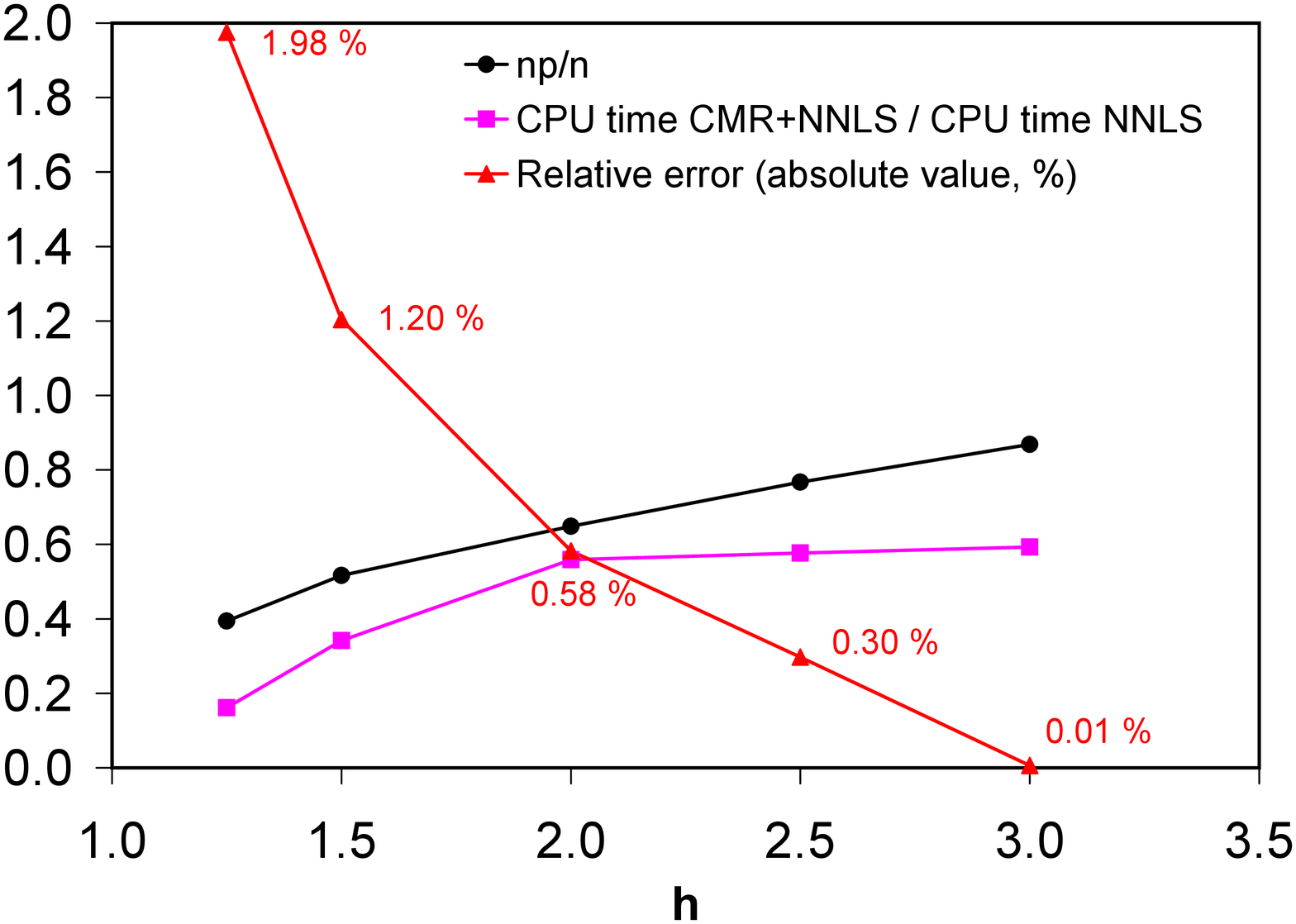}}
\caption{Performance of the CMR+NNLS method with respect to NNLS for
a numerically generated fractal surface with $H=0.7$, depending on
the parameter $h$.}\label{fig9}
\end{figure}

A synthetic diagram illustrating the effect of the parameter $h$ for
the surface with $H=0.7$ and for a single imposed displacement
corresponding to the maximum load in Fig.~\ref{fig8c} is shown in
Fig.~\ref{fig9}. The relative error is rapidly decreasing to values
less than $1\%$ by increasing $h$. The ratio between the number of
points expected to be in contact after the application of the CMR
projection criterion, $n_p$, and the number of points that would be
included by using the classic rigid-body interpenetration check,
$n$, is ranging from 0.4 to 0.8 by increasing $h$ from 1.25 to 3.0.
The ratio between CPU times, on the other hand, tends to an
asymptotic value of 0.6, which implies a saving of 40\% of
computation time as compared to the exact solution, with less than
0.01\% of relative error.

\begin{figure}
\centering
\subfigure[10x]{\includegraphics[width=.48\textwidth,angle=0]{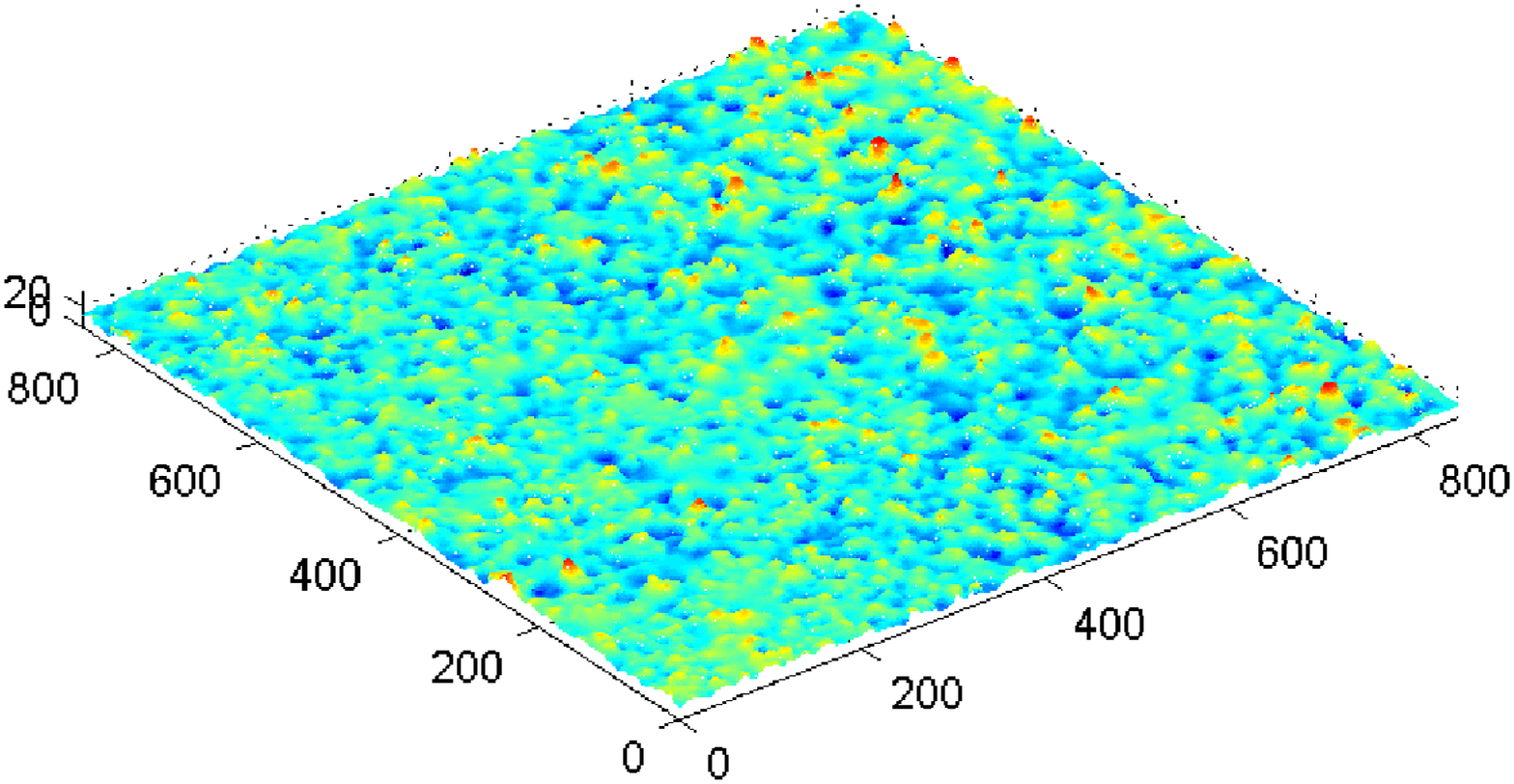}\label{fig10a}}\quad
\subfigure[100x]{\includegraphics[width=.48\textwidth,angle=0]{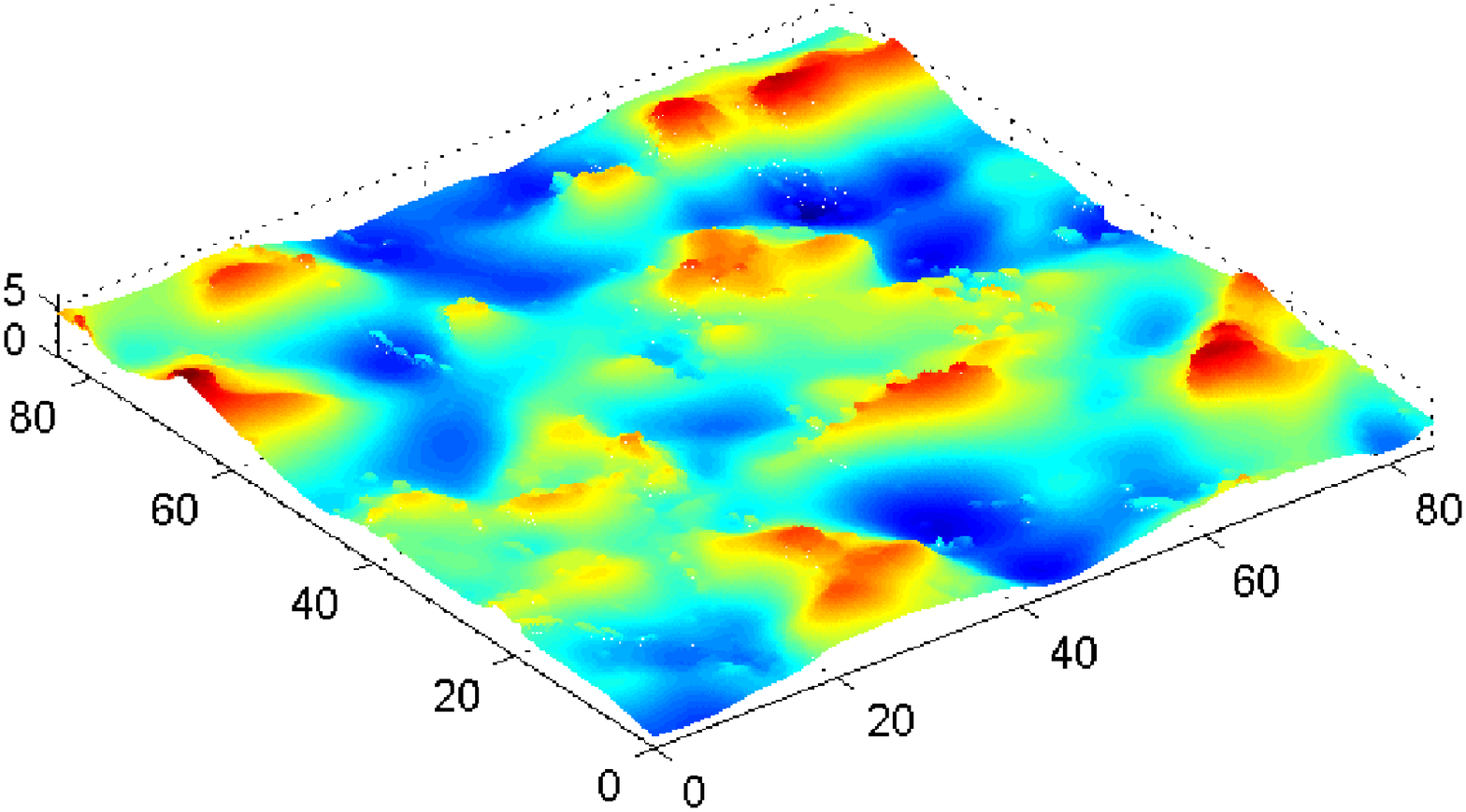}\label{fig10b}}
\caption{Surface of textured silicon solar cells sampled with a
confocal profilometer at two different magnifications (10x and 100x)
obtained by using two different lenses.}\label{fig10}
\end{figure}

We also check the CMR method for warm starting on real surfaces not
displaying the ideal fractal scaling at any length scale, to better
assess possible limits of applicability. As a practical example we
consider the surface of textured silicon solar cells sampled with
two different lenses in order to achieve two different
magnifications (10x and 100x) by using the confocal profilometer
Leica DCM3D, see Fig.~\ref{fig10}. The PSD function of such a
surface sampled with 512 points per side presents a power-law trend
for high frequencies (fine resolutions) and a cut-off to the
power-law at low frequencies (coarse resolutions). In the power-law
regime the surface is characterized by a Hurst exponent $H\cong 0.6$
that can be determined by the slope of the PSD function as
customary.

As a main difference with respect to pre-fractal rough surfaces
generated by the RMD algorithm, the application of the CMR method
requires a filter to downsample the acquired surfaces and extract
their coarser representations. The CMR method is applied to the two
surfaces acquired with 10x and 100x magnifications using $h=1.5$ and
considering a cascade of projections involving coarser
representations of the finest surfaces with 64 and 128 heights per
side. A single contact step corresponding to an imposed far-field
normal displacement equal to $(\xi_{\max}-\xi_{\text{ave}})/5$ is
examined.

The application of the CMR+NNLS method to the surface acquired at
100x leads to very good results in line with those observed for
ideal fractal surfaces. The relative error in the prediction of the
normal load is $-0.4\%$, with a saving of CPU time of $18\%$ as
compared to the direct application of the NNLS algorithm. On the
other hand, the method applied to the surface acquired at 10x leads
to poor results in terms of accuracy with $-98\%$ of relative error
and almost no saving in computation time. This bad performance is
due to the fact that the property of lacunarity of the contact
domain, strictly connected with the self-affine scaling of roughness
due to fractality, does not hold anymore for the surface sampled at
10x due to the cut-off to its power-law PSD. As a consequence, the
CMR method erroneously excludes many possible points from the
initial contact domain suggested by the rigid body interpenetration
check that are actually relevant for contact. Therefore, in
conclusion, the CMR method is efficient for warm starting the NNLS
algorithm, but it should be strictly applied to numerically
generated or real rough surfaces provided that the self-affine
properties of roughness are confirmed by a PSD function of power-law
type.

\begin{figure}
\centering {\includegraphics[width=.55\textwidth,angle=0]{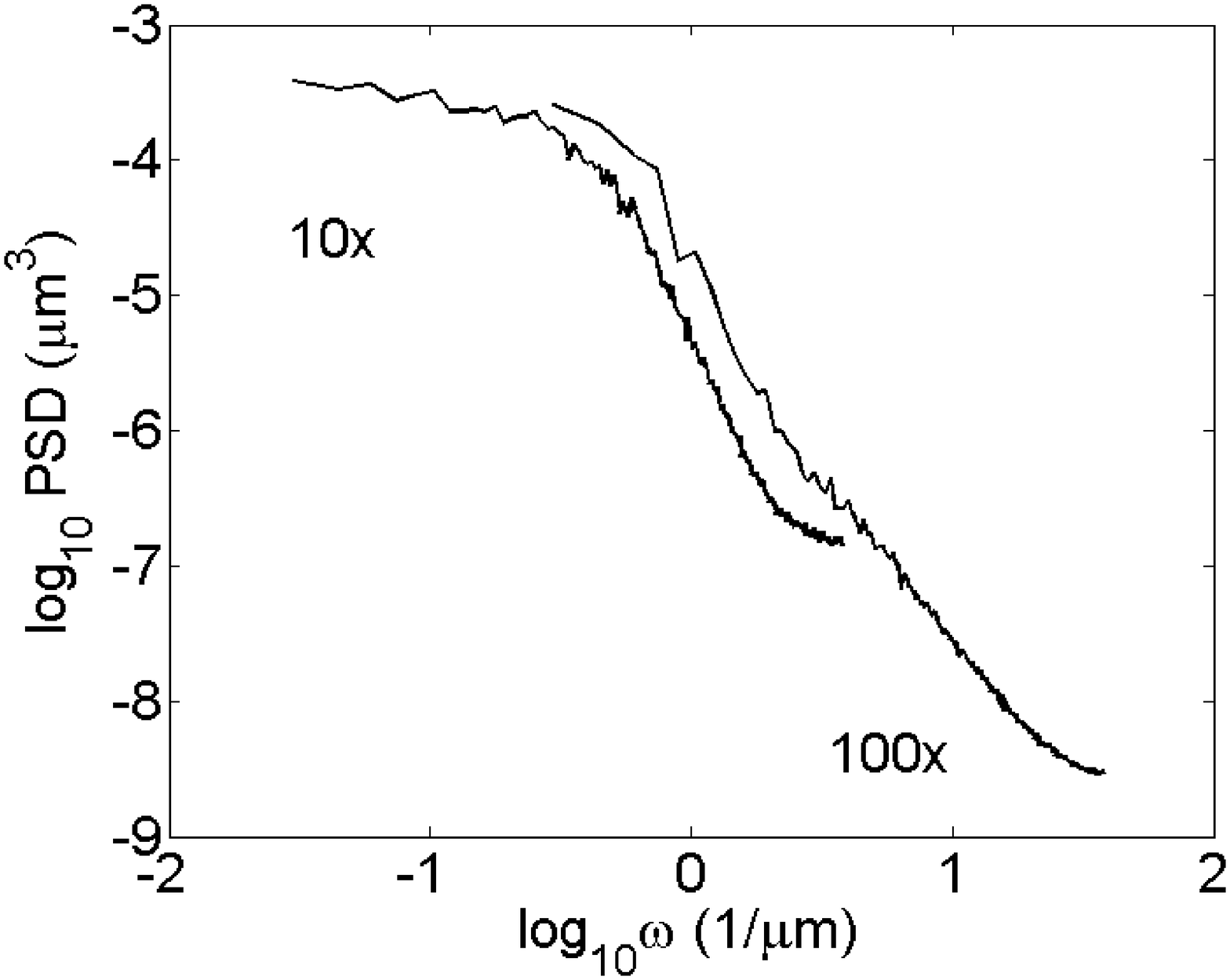}}
\caption{Power spectral density function (PSD) of the two sampled
rough surfaces shown in Fig.~\ref{fig10}.}\label{fig11}
\end{figure}

\section{Conclusion}
This paper has shown how the problem of frictionless normal contact
between rough surfaces within the BEM framework can be solved very
efficiently by exploiting ideas from convex quadratic programming. A
series of efficient optimization algorithms has been proposed and
compared with the traditional Greedy method and constrained CG
algorithm. As the lack of convergence of the Greedy method seems to
be a rare phenomenon, it remains an open question to establish the
conditions on $\bQ$ and $\bar\bu$ for which the algorithm is
guaranteed to converge.

The NNLS algorithm warm started by accelerated gradient projections
was shown at least two orders of magnitude faster than the Greedy
method and 26 times faster than the original constrained CG
algorithm.

Finally, we explored another method for warm starting the
optimization algorithms, this time focusing on a selective reduction
of the size of the initial trial contact domain based on the
multi-resolution properties of roughness. The resulting cascade
multi-resolution (CMR) method allows a further saving of about
$50\%$ of CPU time as compared to NNLS for contact simulations
involving numerically generated fractal surfaces. Relative errors
were found less than $2\%$ for surfaces with $H>0.5$, by using
$h=2$, that was found a good compromise between accuracy and
computation time. Moreover, it has to be remarked that not only the
solution of the finest contact problem is gained by the CMR+NNLS
method with much less CPU time, but also the contact problems
involving all the coarser representations of the finest surface.
These results are particularly important for speeding up intensive
Monte Carlo simulations involving a sequence of contact simulations
for a population of fractal surface with different resolution. So
far, to the best of the authors' knowledge, such extensive
simulations, that are important to determine more reliable trends
from the statistical point of view, have been limited to populations
of 20 to 50 randomly generated surfaces.

In case of real surfaces, a very good performance (less than $2\%$
of error with 3 cascades and at least 18$\%$ of CPU time saved for
one single imposed displacement step) has been demonstrated in case
of power-law PSDs, assuring the self-affine scaling of roughness
which represents the main underlying assumption for the algorithm
applicability. For surfaces with a cut-off to the power-law PSD, on
the other hand, the CMR+NNLS method has given poor results in terms
of accuracy and in any case almost no saving in CPU time as compared
to the pure application of NNLS. Therefore, this warm start method
should be used with care and only in a range where the PSD is of
power-law type.

Finally, we point out that the proposed optimization methods can
also be applied to frictional contact problems by using for instance
the complete BEM formulation as in \citep{pohrt}. Although this
issue is left for further investigation, we expect an even more
significant gain in CPU time by applying the algorithms presented in
this paper instead of other optimization methods, since the size of
the problem is by far significantly increased as compared to the
frictionless case.

\vspace{1em} \addcontentsline{toc}{section}{Acknowledgements}
\noindent\textbf{Acknowledgements} \vspace{1em}

\noindent The research leading to these results has received funding
from the European Research Council under the European Union's
Seventh Framework Programme (FP/2007-2013) / ERC Grant Agreement n.
306622 (ERC Starting Grant ``Multi-field and multi-scale
Computational Approach to Design and Durability of PhotoVoltaic
Modules" - CA2PVM; PI: Prof. M. Paggi).

\end{document}